\documentclass[%
 reprint,
 amsmath,amssymb,
 aps,
]{revtex4-1}

\usepackage{graphicx}
\usepackage{epstopdf}
\usepackage{dcolumn}
\usepackage{bm}
\usepackage{subfigure}

\begin{document}

\preprint{APS/123-QED}

\title{High brightness fully coherent X-ray amplifier seeded by a free-electron laser oscillator}

\author{Kai Li$^{1,2}$}
\author{Jiawei Yan$^{1,2}$}
 \author{Chao Feng$^1$}
 \author{Meng Zhang$^1$}
 \author{Haixiao Deng$^{1,}$}%
 \email{denghaixiao@sinap.ac.cn}
 \affiliation{%
 $^1$Shanghai Institute of Applied Physics, Chinese Academy of Sciences, Shanghai 201800, China\\
 $^2$University of Chinese Academy of Sciences, Beijing 100049, China.
}%

%



\date{\today}

\begin{abstract}
X-ray free-electron laser oscillator (XFELO) is expected to be a cutting edge tool for fully coherent X-ray laser generation, and undulator taper technique is well-known for considerably increasing the efficiency of free-electron lasers (FELs). In order to combine the advantages of these two schemes, FEL amplifier seeded by XFELO is proposed by simply using a chirped electron beam. With the right choice of the beam parameters, the bunch tail is within the gain bandwidth of XFELO, and lase to saturation, which will be served as a seeding for further amplification. Meanwhile, the bunch head which is outside the gain bandwidth of XFELO, is preserved and used in the following FEL amplifier. It is found that the natural ``double-horn'' beam current as well as residual energy chirp from chicane compressor are quite suitable for the new scheme. Inheriting the advantages from XFELO seeding and undulator tapering, it is feasible to generate nearly terawatt level, fully coherent X-ray pulses with unprecedented shot-to-shot stability, which might open up new scientific opportunities in various research fields.
\begin{description}
\item[PACS numbers]
41.60.Cr
\end{description}
\end{abstract}

\pacs{Valid PACS appear here}
\maketitle

\section{\label{sec:level1}Introduction}
The successful operation of self-amplified spontaneous emission (SASE) \cite{kondratenko1980generating,bonifacio1984collective} X-ray free electron lasers (XFEL) \cite{emma2010first,pile2011x,kang2013current,altarelli2006european} around the world offer innovative approach for studying various objects such as nonlinear spectroscopy, diffraction before destruction, time-resolved pump-probe dynamics and coherent scattering \cite{bostedt2016linac}. Starting from initial electron beam density distribution shot noise, SASE is able to produce spatially coherent short wavelength radiation pulse with GW level peak power and around 100 fs pulse duration. In order to increase the resolution of X-ray imaging experiments \cite{chapman2011femtosecond,seibert2011single}, one major goal for the future XFEL is to further enhance peak power and photon flux. Sustainable energy extraction from electron beam to radiation field is guaranteed by resonant condition in FEL process. Saturation occurs for normal planar undulator when electron beams loss too much energy and drop out of resonance with wave. To further extract energy and increase radiation peak power to TW level, the resonant condition should be preserved and undulator parameters ought to be precisely optimized to compensate the electron beam energy loss \cite{kroll1981free}. Lots of efforts have been made to improve the energy extraction efficiency in the last decades \cite{schneidmiller2015optimization,jiao2012modeling}, and the tapering technique has proved to be feasible to improve energy extraction efficiency in long wavelength regions, i.e., microwave radiation and visible light regime \cite{orzechowski1986high,wang2009efficiency}. However, it is predicted that tapered undulator technique is sensitive to shot-to-shot fluctuation and poor temporal coherence of SASE \cite{huang2005free,saldin2006self}, thus effectively enhancing XFEL output power by tapered undulator is still challenging.

Improving the longitudinal coherence of X-ray FEL is of great interest ever since the first light of Linac Coherent Light Source. Modifications to SASE, e.g., self-seeding \cite{geloni2011novel,amann2012demonstration} and slippage-boosted methods \cite{xiang2013purified,mcneil2013transform,wu2012generation}, have been proposed to achieve this goal. Self-seeding takes advantages of monochromator to obtain a coherent seed from SASE which is amplified in downstream undulator. Although self-seeding has been successfully demonstrated to improve the longitudinal coherence, the intrinsic fluctuation of purified seeding is a great challenge. Other methods of increasing the coherent length by speeding up the slippage in undulators, however, will still suffer from large shot-to-shot fluctuations. In addition, for external-seeding technique \cite{yu1991generation,stupakov2009using,deng2013using}, the preservation of temporal coherence and operation stability are still not clear in hard X-ray region. These shortcomings of previous methods are the obstacles to further enhance photon flux by tapering technique.


In order to further improve the XFEL performances by undulator tapering, according to the conventional oscillator-amplifier configuration in laser community, a robust X-ray source which served as a seed is crucial. It is straightforward to take X-ray FEL oscillator (XFELO) output as a high qualified stable seed source and to be amplified in the following FEL amplifier \cite{denghigh}. Utilizing relativistic electron beam as gain medium and crystals as optical cavity mirrors, XFELO starts from spontaneous radiation and generates stable, high repetition rate, fully coherent X-ray pulses, which are promising candidates for seeding FEL \cite{kim2008proposal,dai2012proposal}. In this paper, a novel and simple FEL amplifier seeded by XFELO is proposed by using a chirped electron beam. In this scheme, with the right choice of the beam parameters, only the bunch tail lases inside the XFELO. After over energy modulation by X-ray pulse in the cavity, the degenerate electron beam passes through a chicane delay, and then the bunch head overlaps with the output XFELO pulse in the following FEL amplifier with tapered undulators, where more power is drawn from the electron beam, substantially increases the photon flux. The stable coherent seeds from oscillator ensure high efficiency of tapering technique, and according to the simulation results, nearly terawatt level, ultrahigh brightness, highly stable X-ray pulses can be delivered, which is an ideal light source for scientific users.

Using the parameters of Shanghai Coherent Light Facility (SCLF) as an example \cite{kai2017systematical}, we show that XFELO seeded amplifier is able to significantly enhance FEL peak power, spectral brightness as well as output power stability. It is believed this scheme, which is similar to conventional laser configurations, is quite potential for many future XFEL user facilities. This paper is organized as following: the principles of our scheme is presented in Sec.~\ref{sec:level2}. The simulation results at 5 keV photon energy using a 3 kA flat-top electron beam is illustrated in Sec.~\ref{sec:level3}. The start-to-end simulations on the basis of SCLF are discussed in Sec.~\ref{sec:level4} and the output power stability analysis is shown in Sec.~\ref{sec:level5}. Finally, a brief conclusion is given in Sec.~\ref{sec:level6}.

\section{\label{sec:level2}The principles description}
Unlike conventional lasers whose wavelength is limited by the energy levels of gain materials, XFELO extracts power from relativistic electron beam, thus the output photon energy is unbounded theoretically. Taking advantages of high Bragg reflection crystals demonstrated recently \cite{shvyd2010high}, the X-ray radiation, which starts from spontaneous radiation, is trapped inside the resonator and oscillates back and forth. During each round trip, X-ray pulse overlaps with electron beam and obtains energy insides the undulator. In order to ensure the growth of light power, undulator should be long enough to maintain sufficient single pass gain, i.e.
\begin{equation}
R(1+G)>1,
\label{gain}
\end{equation}
where $R$ is the total reflectivity of the crystal mirrors, and $G$ is the single pass gain. The X-ray radiation is amplified over and over again and experiences exponential growth until saturation. In the saturation regime, electron bunch is over modulated by the high power X-ray which leads to gain degradation. When the round trip net gain equals to one, the XFELO reaches equilibrium and produces stable output X-ray pulses.

\begin{figure}
\includegraphics[width=8cm]{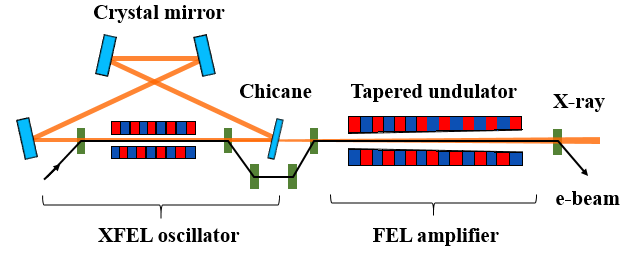}
\caption{\label{schematic}The schematic of XFELO seeded FEL.}
\end{figure}

Similar to traditional laser, with carefully optimized electron beam quality, undulator length and cavity output coupling, XFELO is able to deliver stable, high repetition rate, fully coherent X-ray pulses \cite{li2017simplified,lindberg2011performance}. The peak power is typically several MW for low peak current electron beam ($\thicksim$10A) in the early XFELO proposals \cite{kim2008proposal,dai2012proposal,lindberg2011performance,borland2007configuration}, and due to the narrow spectral acceptance of the Bragg crystal \cite{shvyd2012spatiotemporal}, typically tens of meV, the average brightness for XFELO output is predicted to be 2 orders of magnitudes higher than SASE. With the development of superconducting accelerator technology, several kA peak current Linac-based high repetition rate X-ray FELs are under construction \cite{pierini2017fabrication,burrill2017vertical}, which is suitable for high current XFELO operation. For a typical FEL oscillator, the maximum fractional energy extracted from electron beam is inverse proportion to $1/N_u$, where $N_u$ is the number of undulator periods. For a given single pass gain, high peak current means short undulator length, which is expected to generate X-ray pulse with peak power nearly GW level.  For further enhancing the peak power, Fig.~\ref{schematic} shows the schematic diagram of high peak current XFELO supplies a seed laser for subsequent tapered FEL amplifier. A chicane is applied to displace the electron beam horizontally to bypass the crystal mirror and to match the delay of X-ray pulse by crystal. The details of configuration and properties are investigated in the following.
\subsection{The drawbacks of cavity detune}
Since a high peak power FEL amplifier is pursued, only high peak current electron beam is considered in this paper. In addition, large amount of electrons may drop out of the ``ponderomotive bucket'' after interactions with intense X-ray electromagnetic field in the cavity, which will degrade the tapered FEL efficiency. Thus the beam energy spread degradation in the previous XFELO must be avoid, and the faint X-ray pulse inside the cavity is preferred as long as it dominates over the electron beam spontaneous radiation. For example, an XFELO output with few MW peak power is sufficient as a seeding to overcome the shot noise from a typical 1 kA peak current beam. To control the XFELO power at a certain level, the simple and straightforward options are fine tuning of X-ray cavity, e.g., by increasing the output coupling ratio or by detuning the cavity length.

\begin{figure} 
\includegraphics[width=9cm]{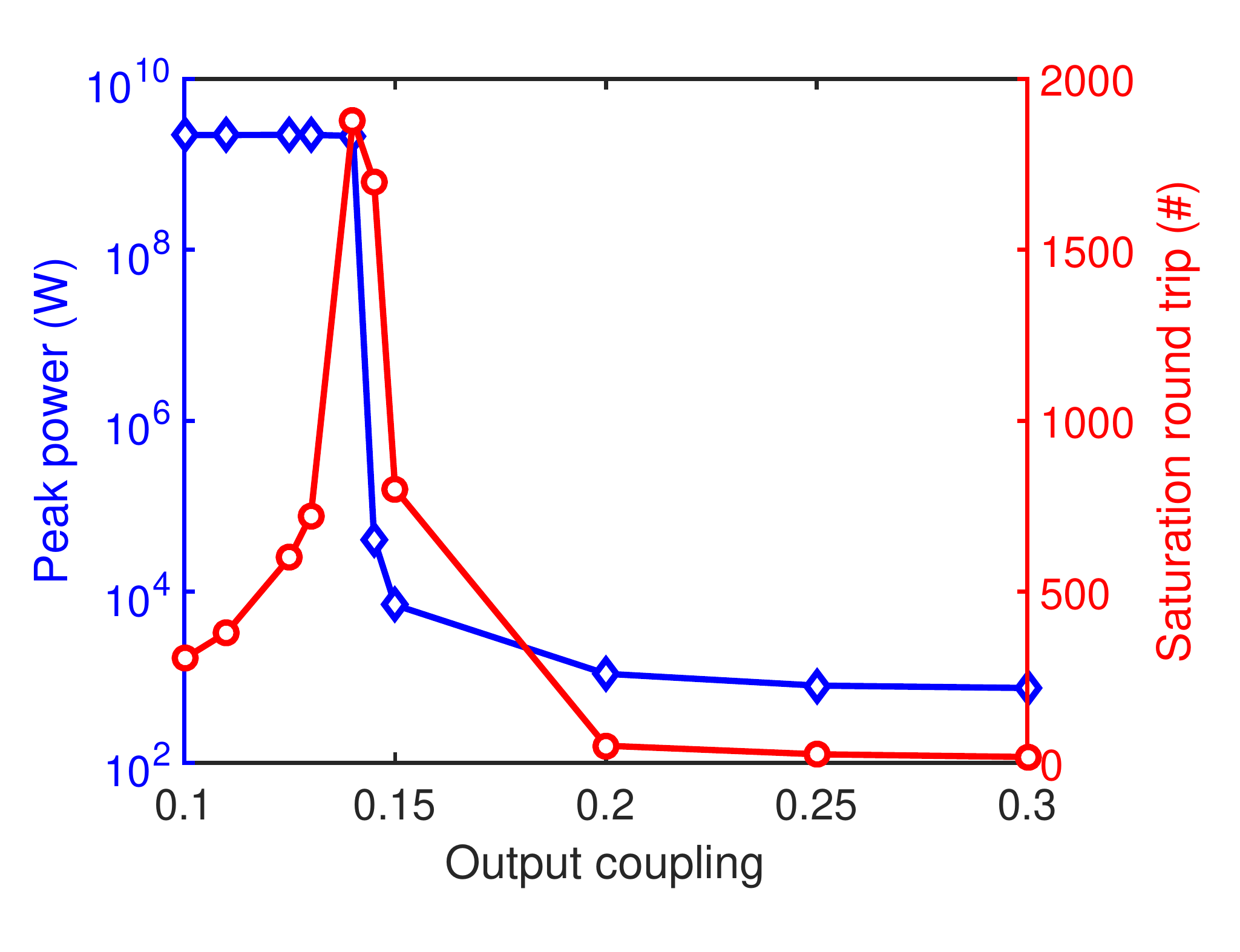}
\caption{\label{reflectivity}A typical 1 kA peak current XFELO output peak power and saturation round trip number as a function of cavity output coupling.}
\end{figure}
\begin{figure} 
\includegraphics[width=8cm]{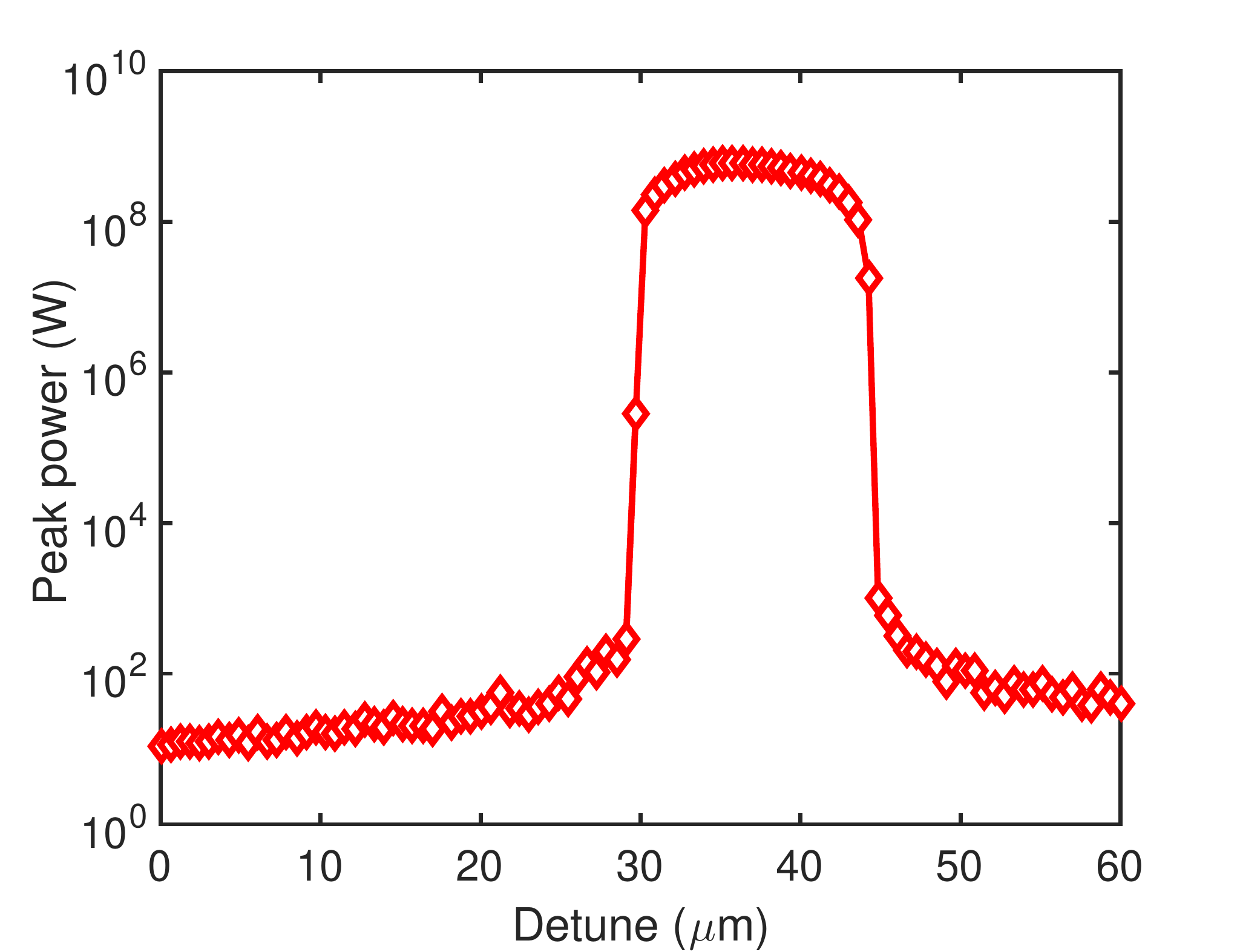}
\caption{\label{detune}The XFELO output peak power at different cavity detune.}
\end{figure}

According to the simulation results, however, aforementioned two approaches are proved to be infeasible. As the output coupling rises, the decline of net reflectivity causes dissatisfactory of Eq.~(\ref{gain}), and the XFELO output peak power decreases rapidly. The steep fall indicates that generating MW level X-ray signal as a seeding pulse by increasing cavity output coupling is unpractical. Since the peak power would either remains shot noise or grows to GW level, it is hard to operate at an unstable status. Note that changing mirrors reflectivity finely by varying crystal thickness is challenging in practical. The round trips required for XFELO saturation is also plotted in Fig.~\ref{reflectivity}, which shows that much more oscillation passes are required to reach an equilibrium status around the critical output coupling. The fine detuning of the cavity length is also infeasible, according to Fig.~\ref{detune}, due to the dramatically growth and decline of output peak power from the optimum synchronization. In order to obtain and maintain a suitable MW output peak power, cavity length control system with nm resolution is required. And any jitter and vibration of cavity mirrors will lead large output power fluctuation and undermines the following tapered FEL amplifier.

\subsection{The chirped-beam XFELO}
\begin{figure} 
\includegraphics[width=8cm]{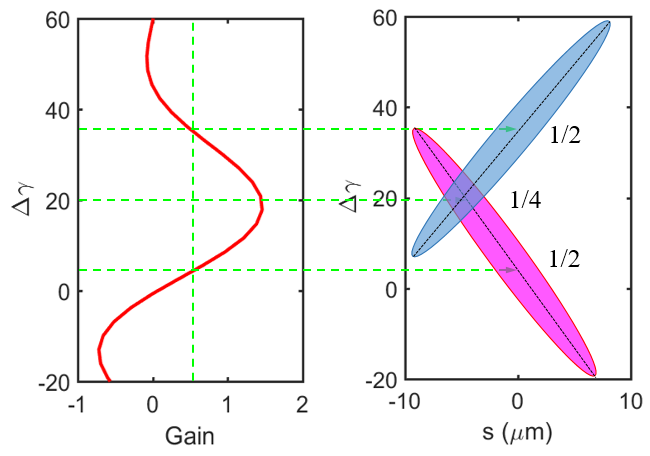}
\caption{\label{fig_chirp}The schematic of 3 kA peak current XFELO single pass gain (red solid line) and optimized chirped-beam. Assuming that 50\% gain is necessary, and the $1/4$ bunch tail point corresponds to the maximum single pass gain, while the central point relates to 50\% single pass gain. Thus only bunch tail gets enough gain to compensate the round trip net loss, while bunch head does not lase inside the XFELO. The blue and pink ellipses represent the positive and negative chirped-beam, respectively.}
\end{figure}

According to the previous results and discussions, the alternative strategies of using XFELO output as a seeding is the subsequent FEL amplifier would be using twin bunch lasing \cite{weilun2017xfelo} or recently developed fresh-slice technique \cite{lutman2016fresh}, which however require relatively complicated machine setup and additional accelerator hardware components. In this paper, however, a simpler chirped-beam XFELO scheme is proposed for preserving energy spread of part of electron beam while getting enough seed radiation from the other part. Schematic for optimization of a 3 kA peak current chirped-beam for XFELO is demonstrated in Fig.~\ref{fig_chirp}. Assuming that single pass gain of 50\% is needed for compensating round trip loss and output coupling. In order to guarantee only the bunch tail lases in XFELO, the electron energy of the beam center is set corresponding to 50\% single pass gain, while the 1/4 point at the bunch tail obtains the maximum single pass gain. Thus the optimized electron energy chirp is，
\begin{equation}
\alpha=\frac{\Delta \gamma / \gamma_0}{\Delta l / L_b}=4~\frac{\gamma_{1/2}-\gamma_{1/4}}{\gamma_0},
\label{eq_chirp}
\end{equation}
where $\gamma_0$ is the resonant electron energy measured in rest mass units, $L_b$ is the length of electron beam and $\Delta l$ is the length deviation corresponding to $\Delta \gamma$ energy detune, and $\gamma_{1/4}$, $\gamma_{1/2}$ represent the electron energy at the 1/4 and 1/2 points. Note that an electron beam with both positive chirp (blue ellipse) and negative chirp (pink ellipse) can be used for the XFELO seeded FEL amplifier. The effect of linear chirp on FEL gain is equivalent to the linear tapered undulator, and thus the beam energy chirp could be compensated with the linear undulator taper and thus optimize the following amplification.

\subsection{Tapered FEL amplifier}
Tapered undulator technique is well-known for the capability of maintain resonant condition after FEL saturation, thus improving the energy extraction efficiency. In practice, however, it is hard to generate ultra-high peak power X-ray pulses as estimated by simulations. As far as we know, one of the main reasons is lack of suitable seed radiation from previous undulators. 

On the one hand, SASE starts from electron initial random density fluctuations, and is known to possess poorly longitudinal coherence, thus unstable generation of parasitic radiation with different frequency from the signal would occur \cite{kroll1981free}, which causes the degeneration of tapered amplifier performance. On the other hand, self-seeding scheme is able to generate X-ray pulses with narrow bandwidth, however, nearly twice as much as undulator periods are required comparing to SASE, and more undulators might be necessary for further improving the pulse energy by tapering technique. In addition, the narrow acceptance of Bragg crystal diffraction for a broad bandwidth SASE from upstream undulators leads to large shot-to-shot self-seeding pulse energy fluctuations, and matching the tapered undulators configurations for each individual shot is impossible, thus leads to even larger output pulse energy fluctuations. Furthermore, the relatively small seed signal after self-seeding, which means small ``ponderomotive bucket'' height, combines with large energy spread might cause portions of electron escaping from bucket and further degradation of tapered FEL amplifier. 
\begin{figure*} 
\centering
\subfigure{\includegraphics[width=5cm]{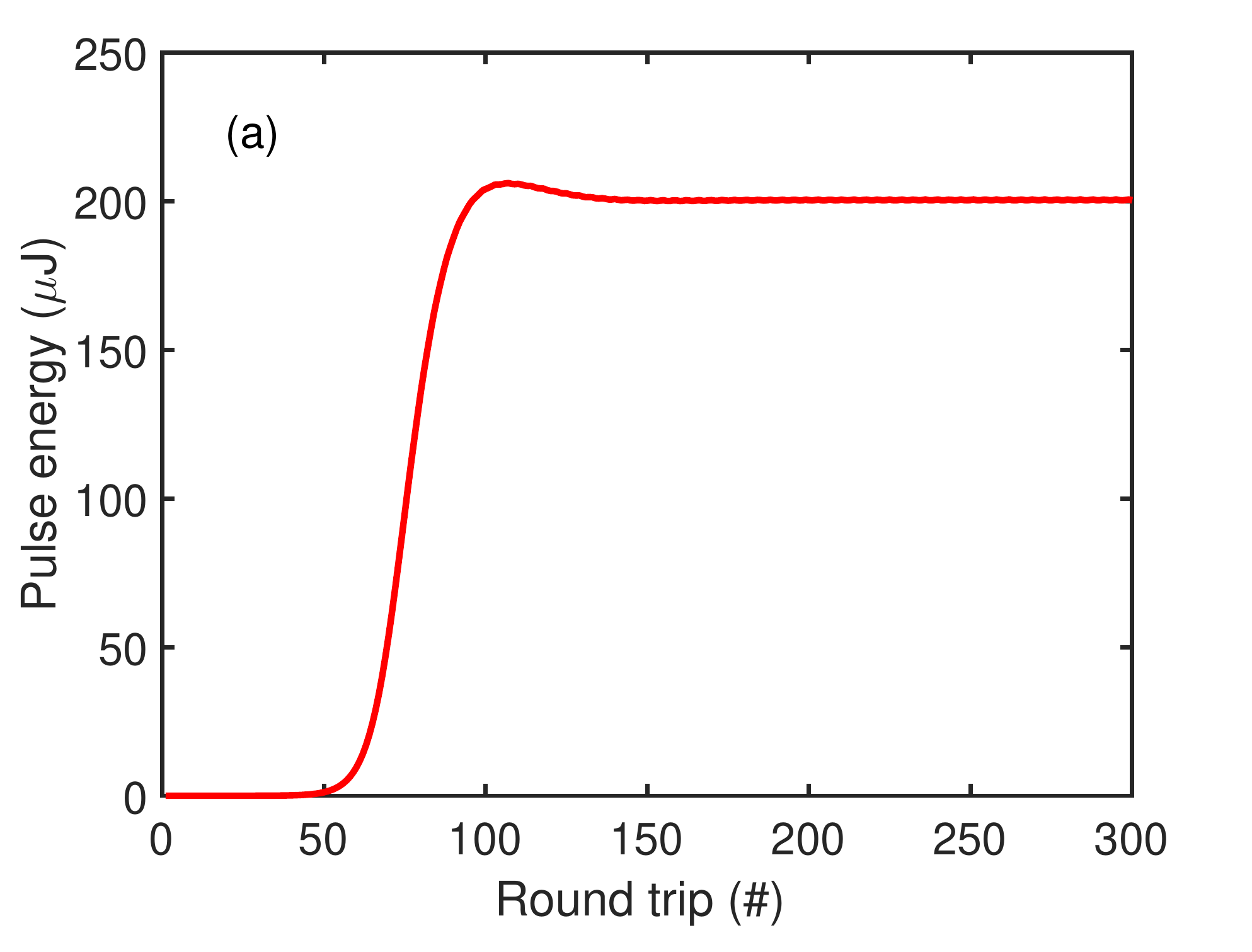}}
\subfigure{\includegraphics[width=5.2cm]{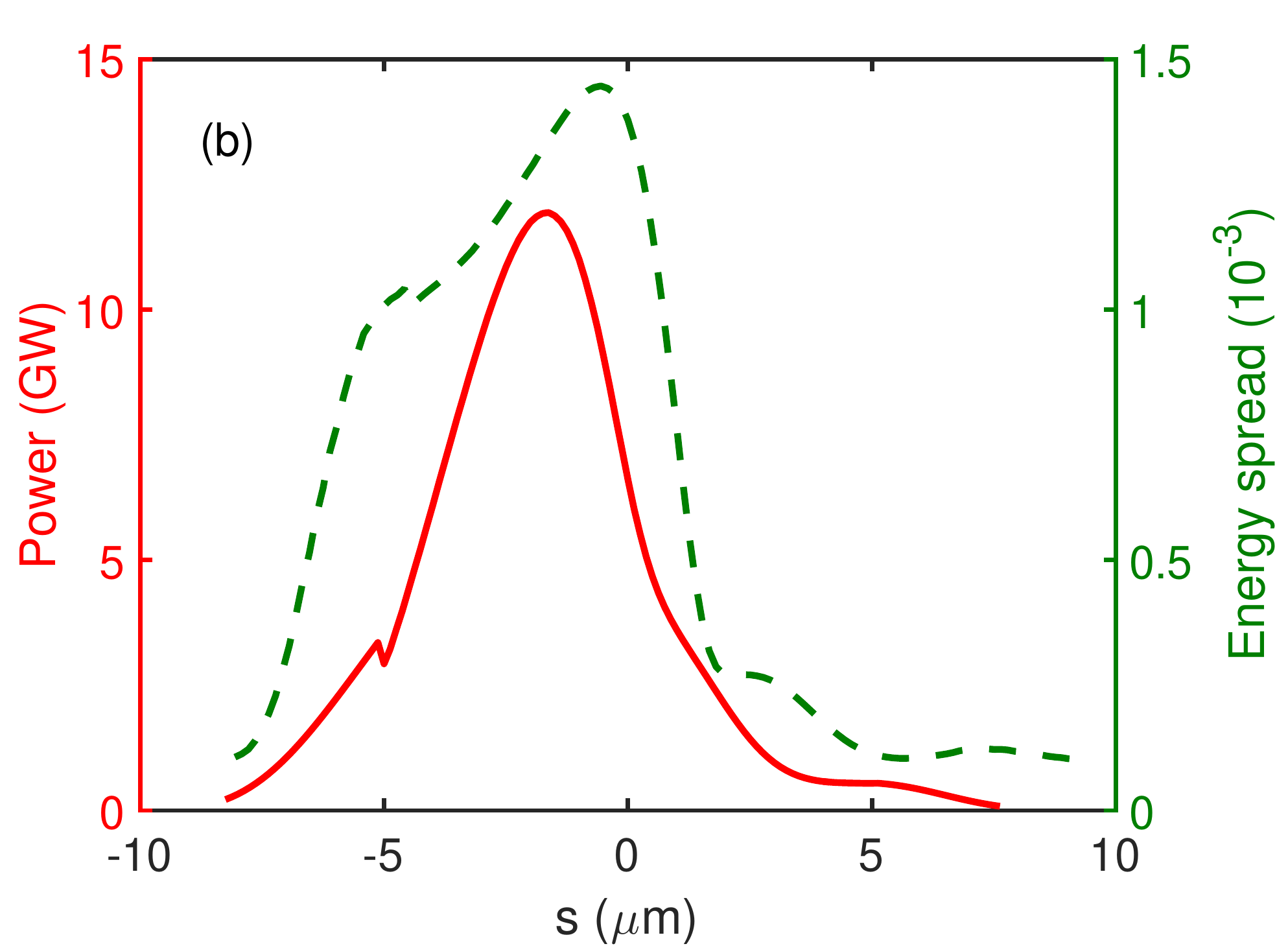}}
\subfigure{\includegraphics[width=5cm]{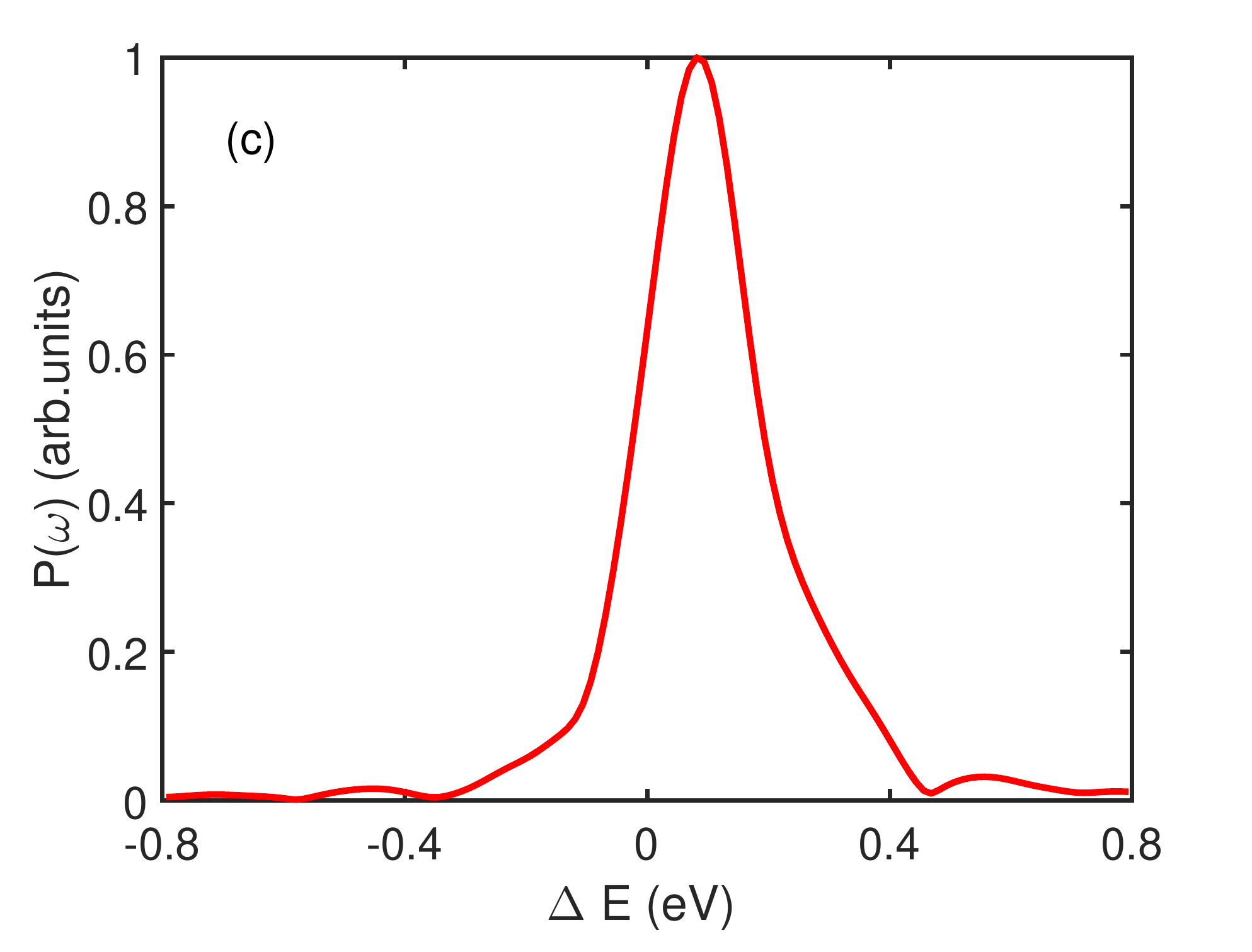}}
\subfigure{\includegraphics[width=5cm]{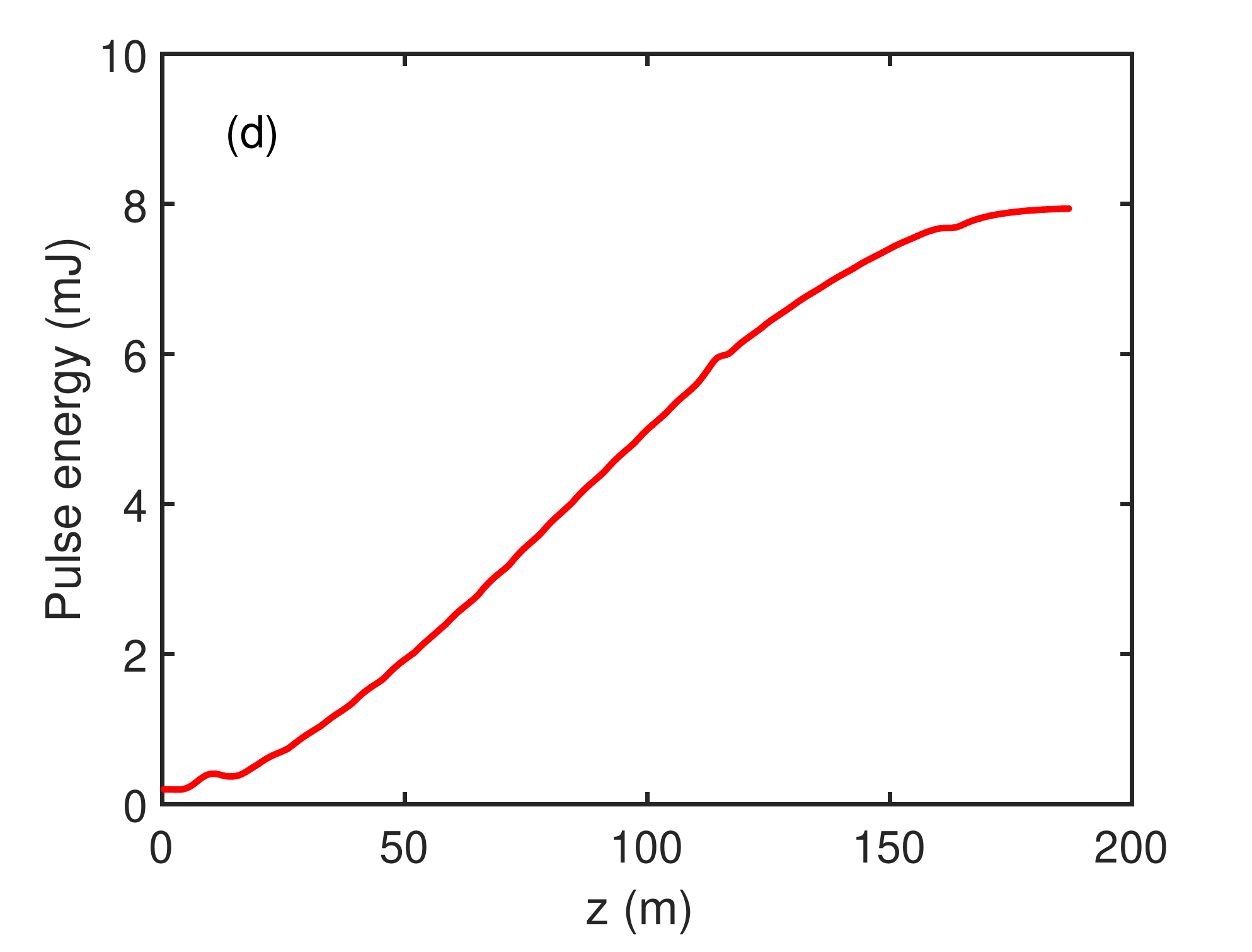}}
\subfigure{\includegraphics[width=5.1cm]{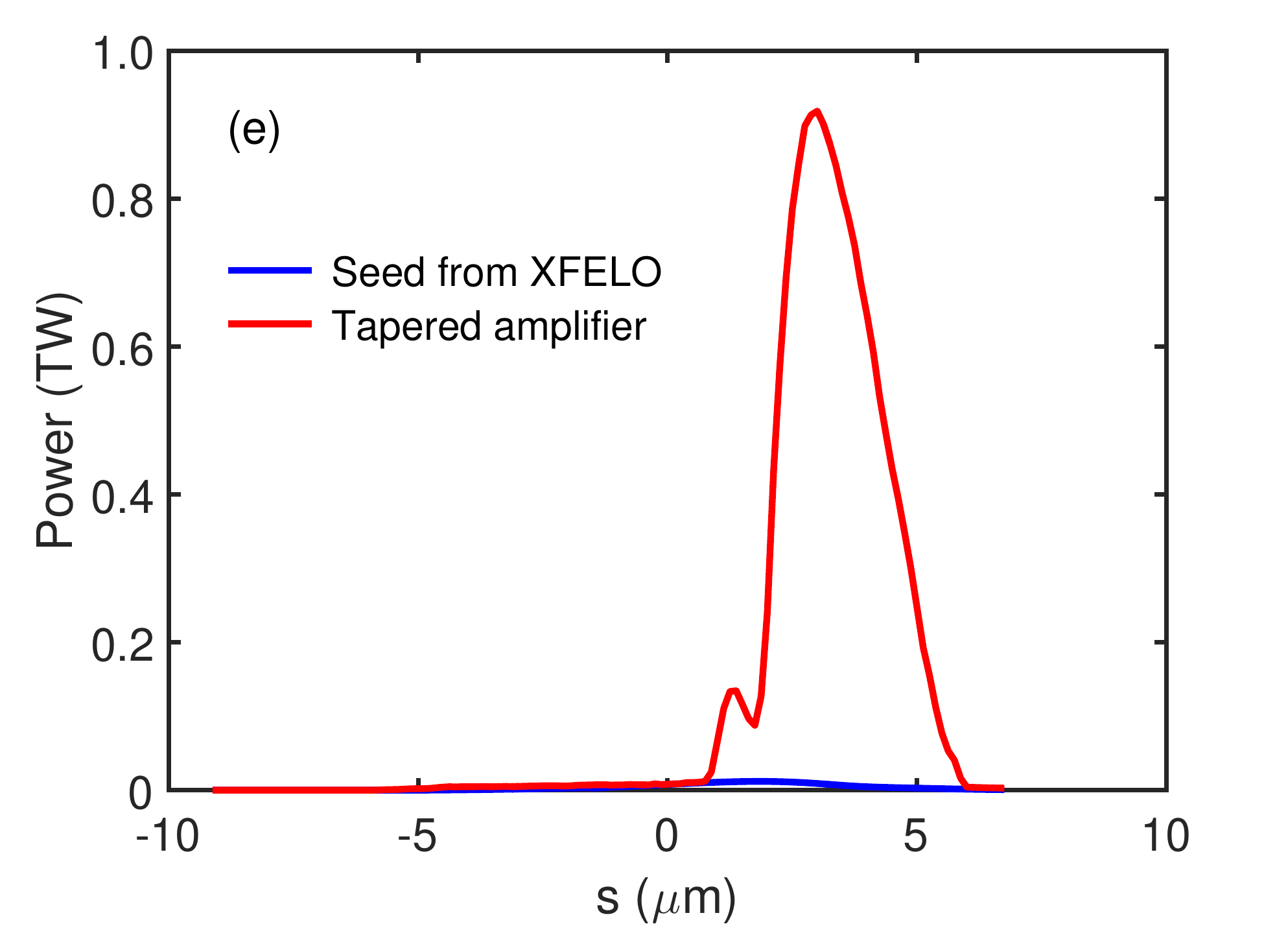}}
\subfigure{\includegraphics[width=5cm]{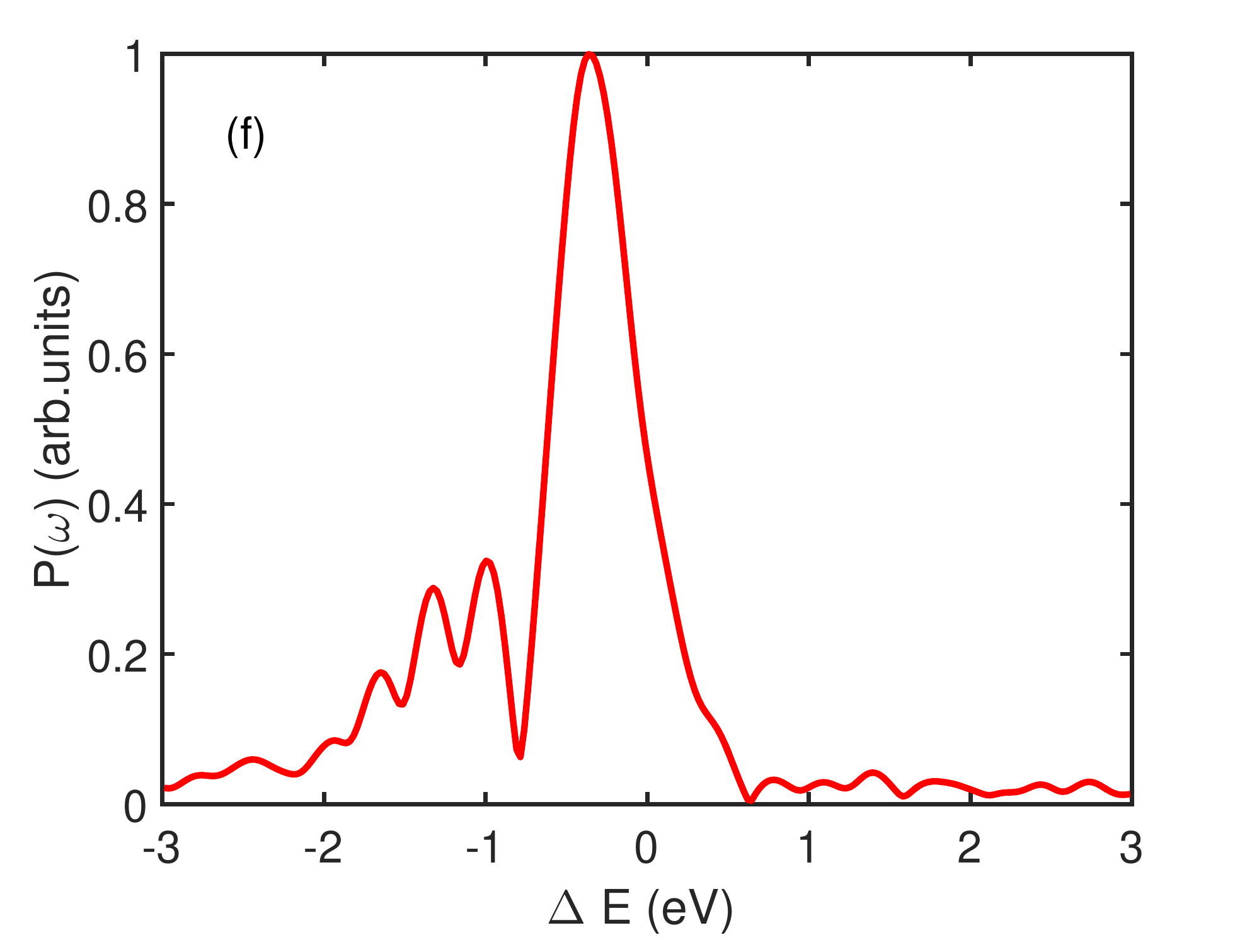}}
\caption{\label{fig_gaussian}The performance of 3 kA peak current flat-top electron beam for XFELO with 18\% crystal mirror output coupling, and for tapered FEL amplifier. The top row of panel shows (a) output pulse energy, (b) power profile (red solid line) and beam energy spread (green dashed line), (c) spectrum for XFELO, while the bottom row of panel (d),(e) and (f) display the corresponding results for tapered FEL amplifier. }
\end{figure*}
\begin{figure*} 
\centering
\subfigure{\includegraphics[width=6cm]{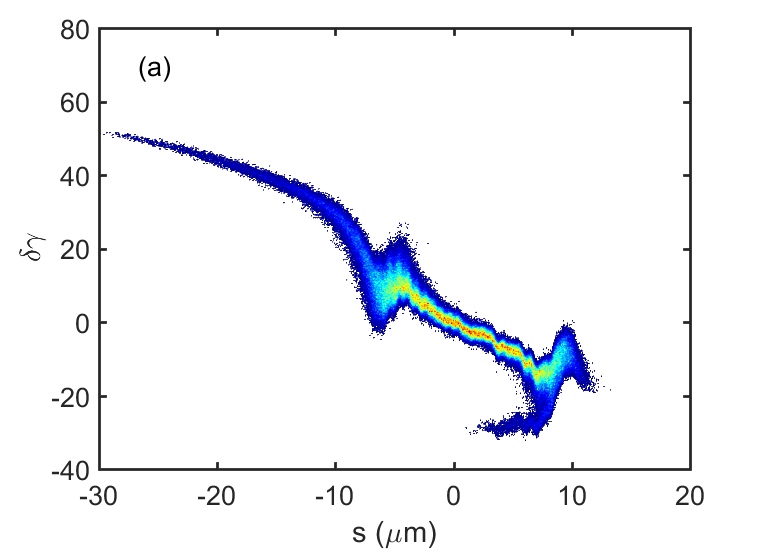}}
\subfigure{\includegraphics[width=6cm]{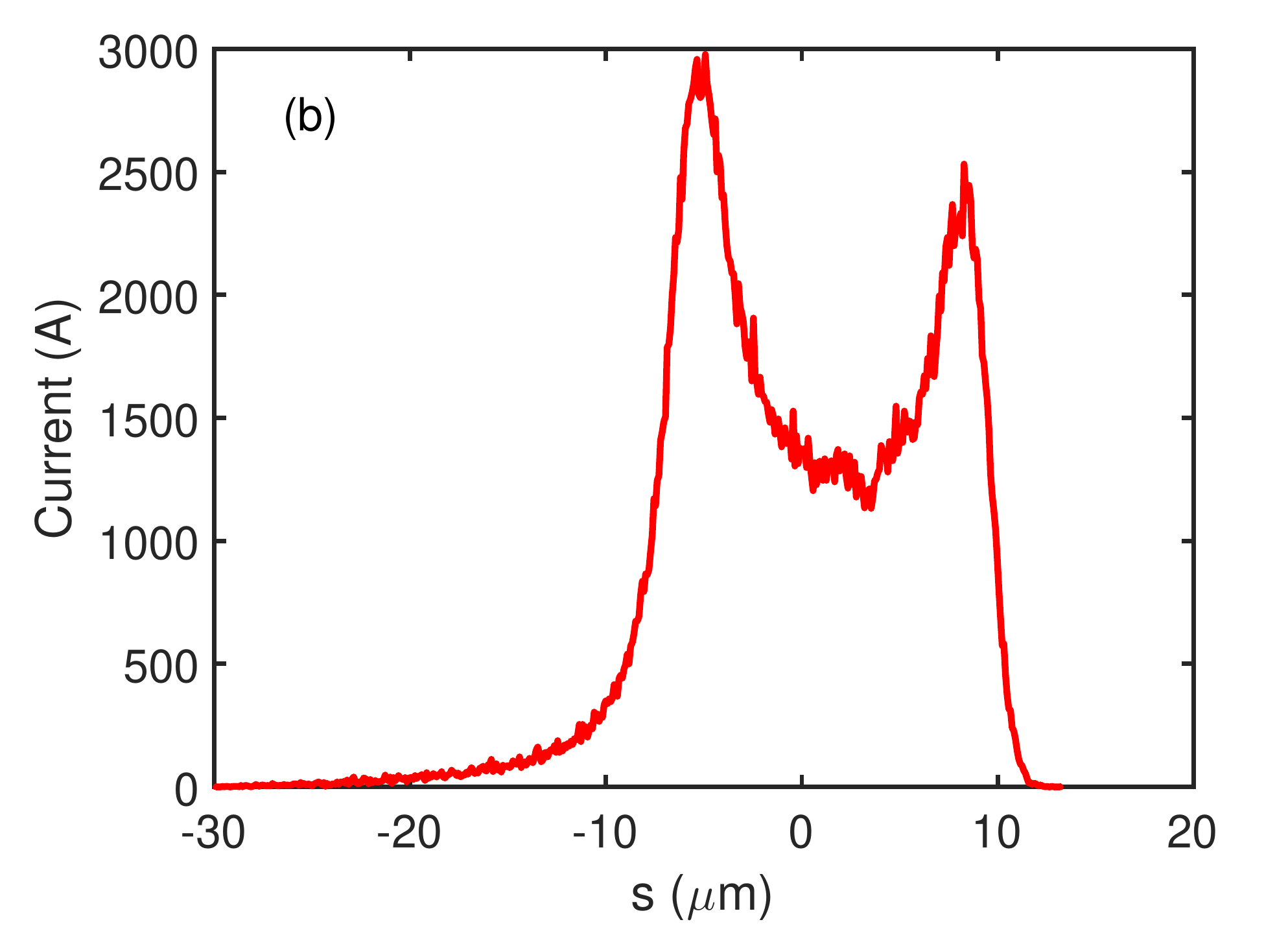}}
\caption{\label{fig_phase}The electrons distribution in phase space (a) and the corresponding current profile (b).}
\end{figure*}

For the chirped-beam XFELO seeded FEL proposed in this paper, a fully coherent, high peak power stable X-ray seed radiation from XFELO is benefit for the following tapered amplifier. And the degenerate portions of lasing electron beam is replaced by another fresh parts of bunch, thus ensures a small energy spread. To eliminate the influence of phase mismatching of tapered FEL amplifier, the undulators without break sections are employed in simulations. The time-dependent results in the following sections proved that it is feasible to generate high brightness X-ray pulses with considerable stability.

\section{\label{sec:level3}Time-dependent Simulation}
To illustrate the main principles of XFELO seeded FEL, the normal parameters of SCLF are employed. SCLF is the first hard X-ray FEL facility in China which is now under conceptual design, and the major parameters of electron beam and undulators are summarized in Table.~\ref{tab:parameter}. SCLF takes advantages of superconducting accelerator technique which is capable of delivering nearly 1 MHz high repetition rate relativistic electron bunches. The combination of high brightness electron beam and 200 m length undulators are suitable for XFELO seeded FEL amplifier. The simulations are specially carried out at 5 keV X-ray photons, due to the production of high peak power and short temporal duration FEL radiation at that wavelength is of great interests for single-particle imaging and coherent diffraction imaging experiments.
\begin{table}
\caption{\label{tab:parameter}The main parameters of SCLF.}
\begin{ruledtabular}
\begin{tabular}{lcr}
Parameter & Value & Unit\\
\colrule
Beam energy & 8 & GeV\\
Slice relative energy spread  & 0.01 & \% \\
Normalized emittance  & 0.4 & $\mu$m-rad\\
Repetition rate & 1 & MHz\\
Peak current  & 3 & kA\\
Bunch charge  & 100 & pC \\
Undulator period  & 26 & mm\\
Total undulator length  & 200 & m \\
\end{tabular}
\end{ruledtabular}
\end{table}

For the benefits of improving tapering efficiency, high intensity ($\sim$3 kA) flat-top current with bunch charge equals to 100 pC electron beam is used. The short electron bunch temporal duration corresponds to a broad spectral bandwidth (especially for XFELO seeded FEL in which only half of electron beam is lasing), and in order to avoid excessive cavity loss due to crystal mirror narrow bandwidth acceptance \cite{shvyd2012spatiotemporal}, the (1 1 1) atom plane of diamond crystal with incident angle $\theta=36.9^\circ$ is chosen to obtain a 324 meV Darwin bandwidth. A 300 m round trip length optical cavity and one module of undulator with 200 periods is used for matching the electron beam repetition rate and providing enough XFELO single pass gain.

The performances of XFELO seeded FEL are simulated by the combination of GENESIS \cite{reiche1999genesis} and OPC \cite{van2009time}. The evolution of XFELO radiation pulse energy is shown in Fig.~\ref{fig_gaussian}(a), which experiences an exponential growth as expected and generates 206 $\mu$J X-ray in the saturation regime. With the optimized electron energy chirp $\alpha=6.1\times10^{-3}$, only the bunch tail, which provides enough FEL gain, contributes to effective FEL interaction as well as induced significant energy spread enhancement. Due to the spectral purification of crystal mirror Bragg diffraction, the output spectral bandwidth is 212.5 meV (FWHM).
\begin{figure*} 
\centering
\subfigure{\includegraphics[width=5cm]{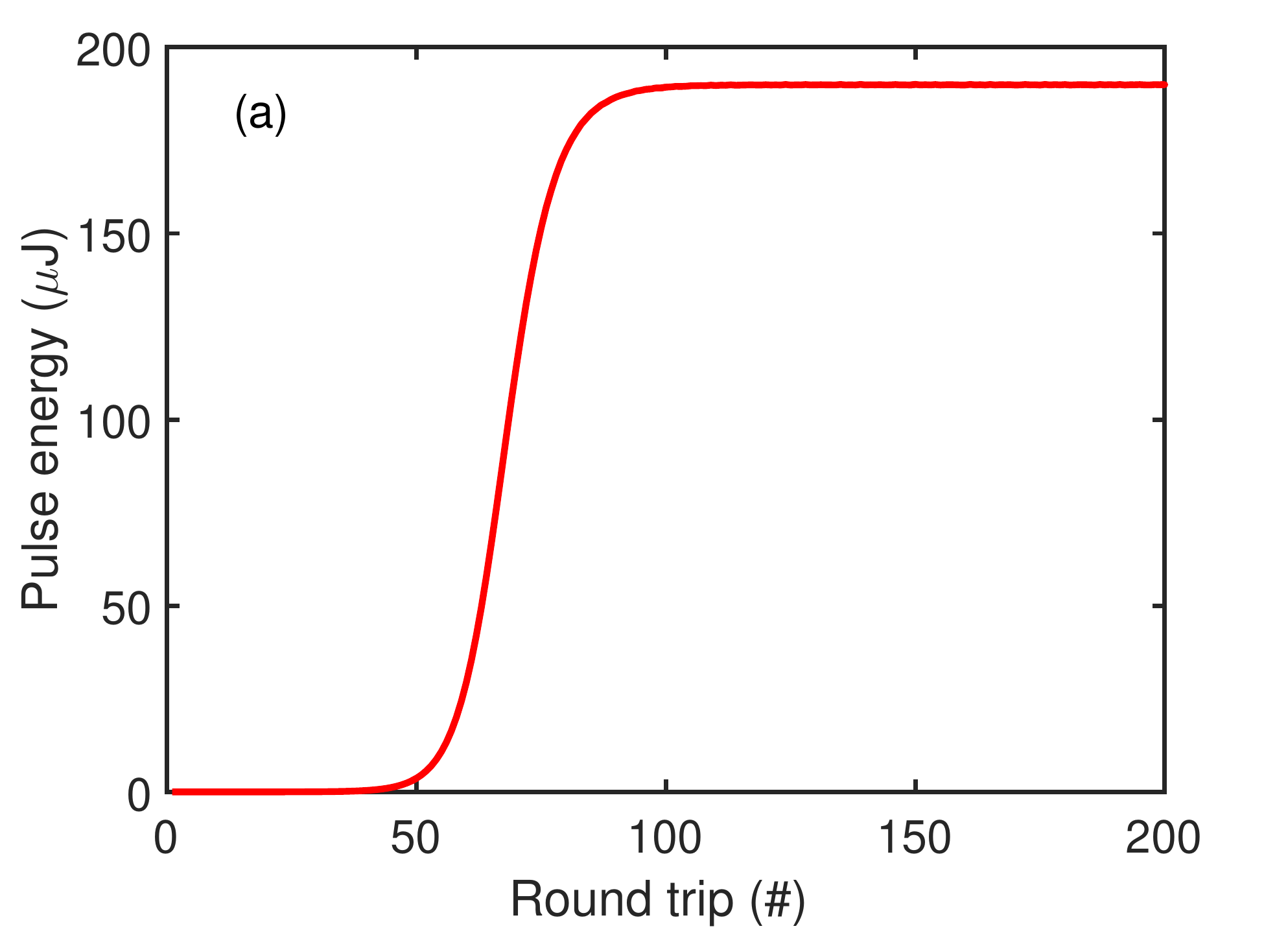}}
\subfigure{\includegraphics[width=5.1cm]{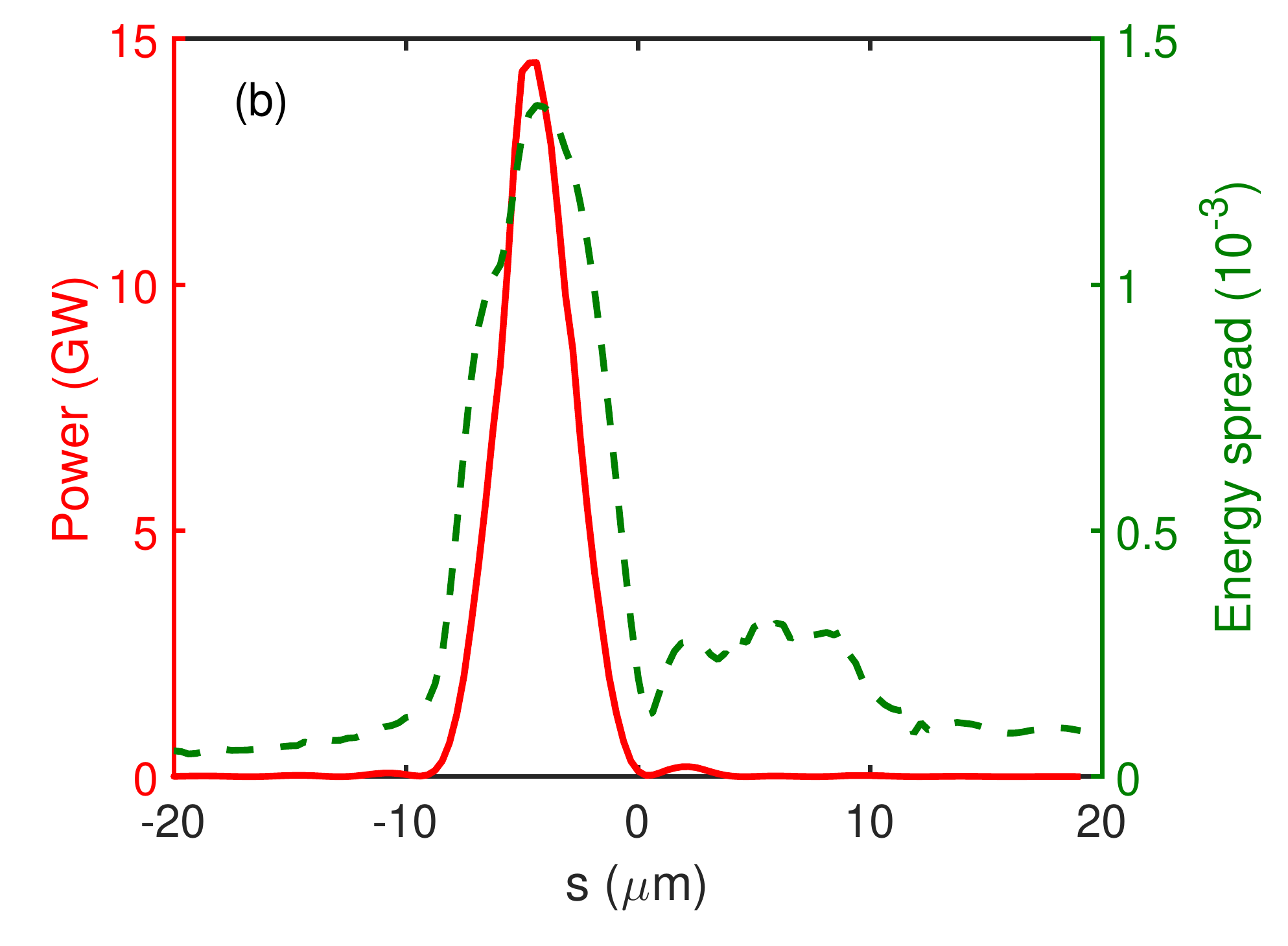}}
\subfigure{\includegraphics[width=5cm]{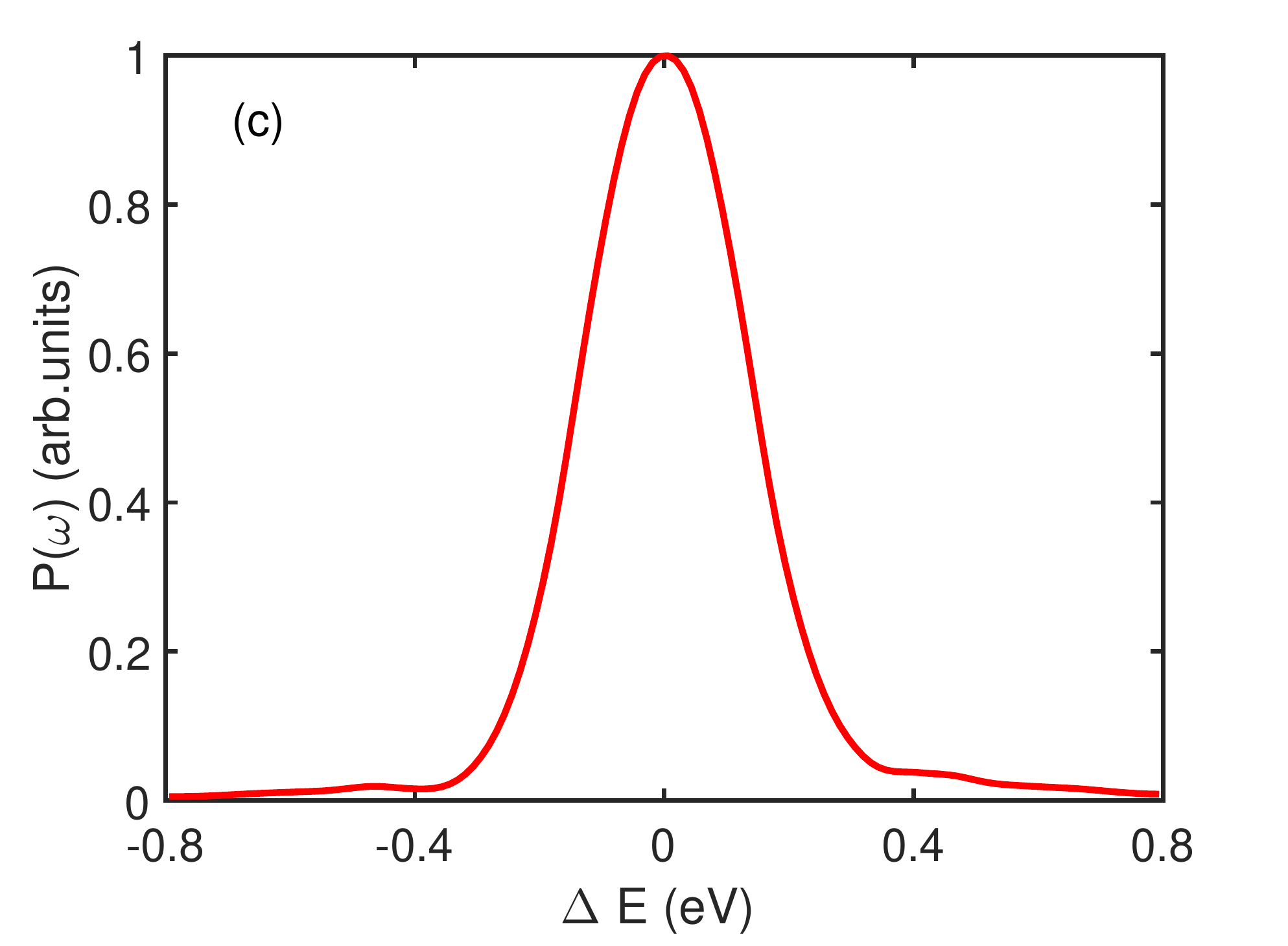}}
\subfigure{\includegraphics[width=5cm]{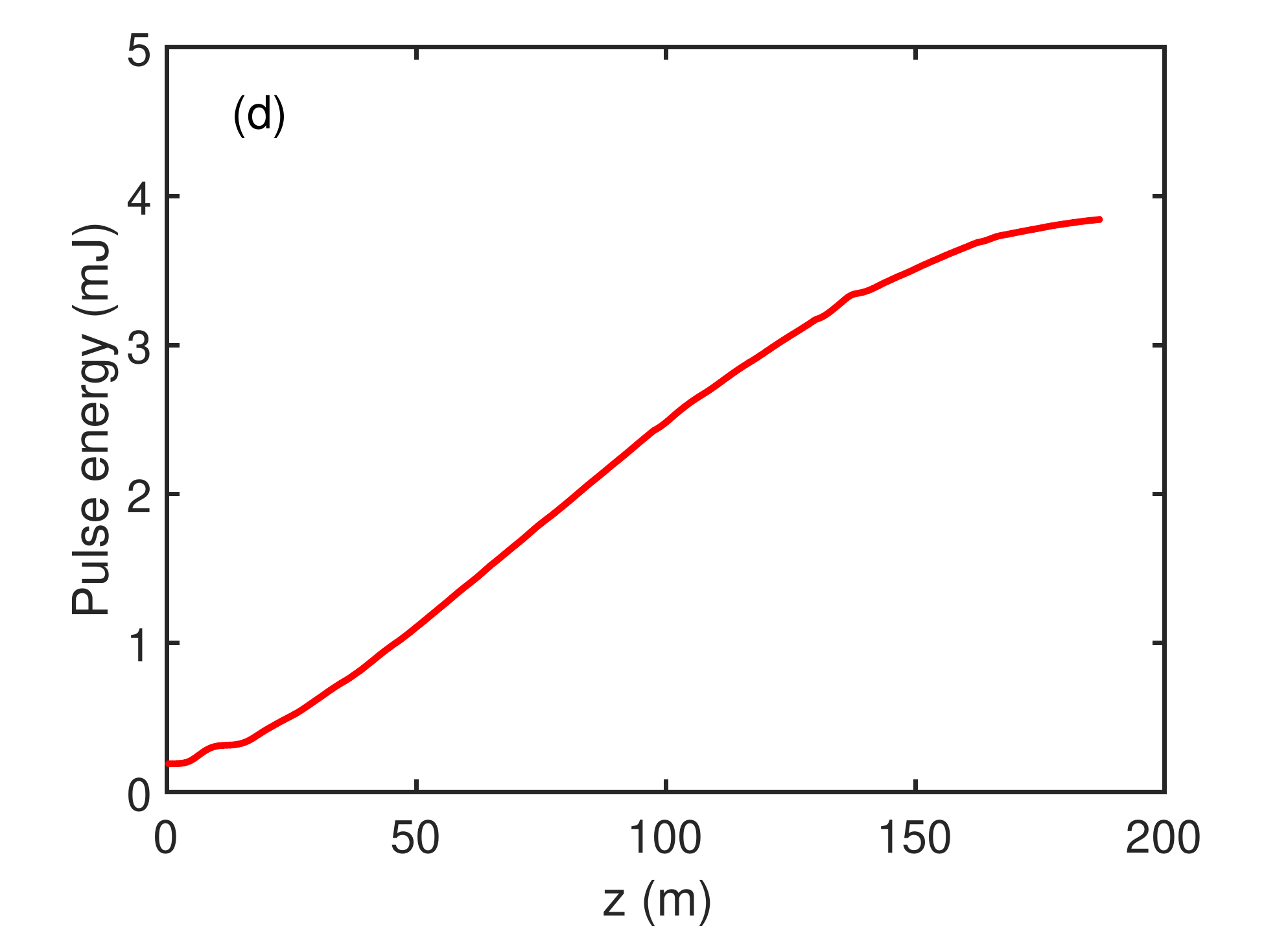}}
\subfigure{\includegraphics[width=5cm]{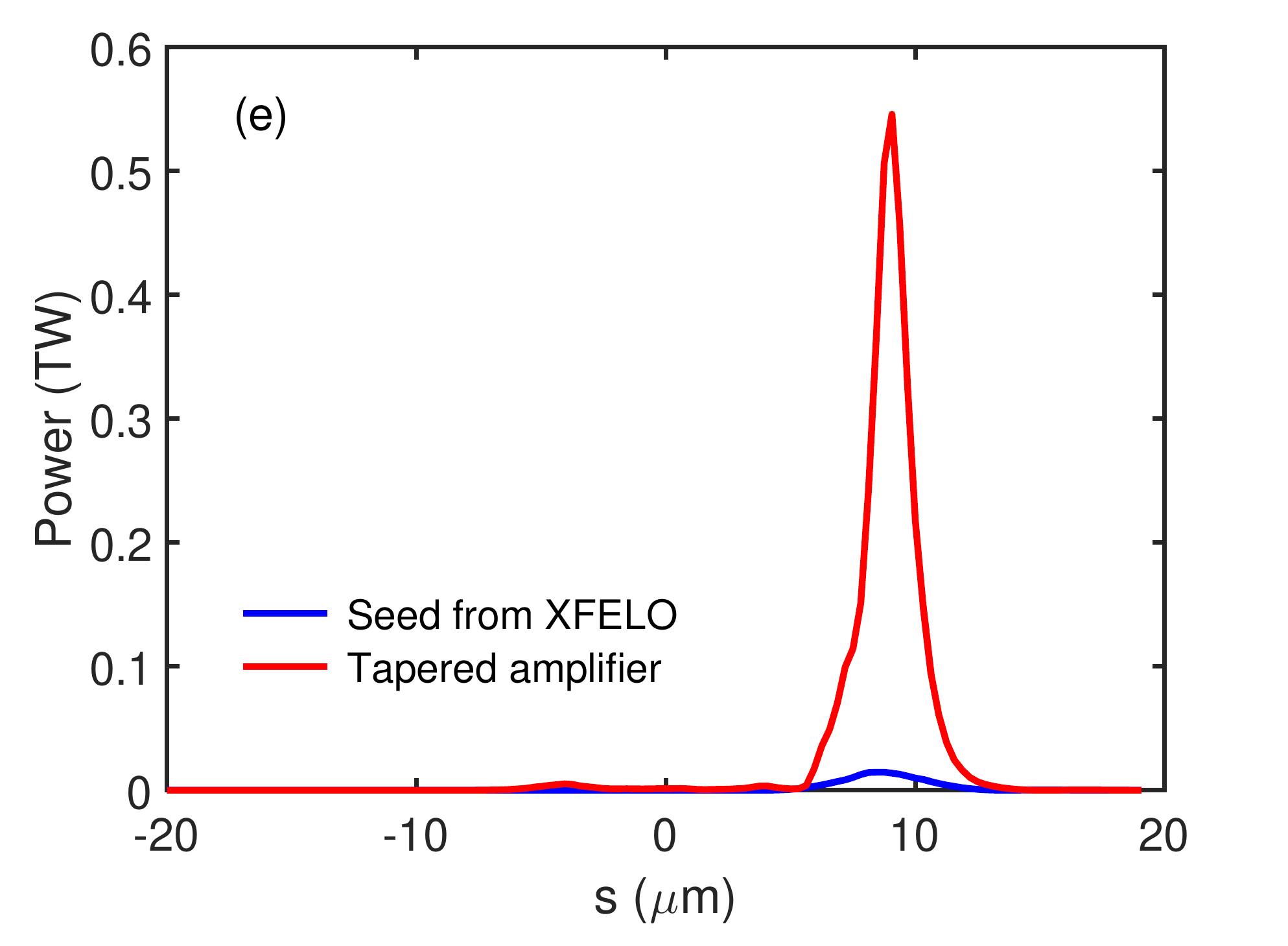}}
\subfigure{\includegraphics[width=5cm]{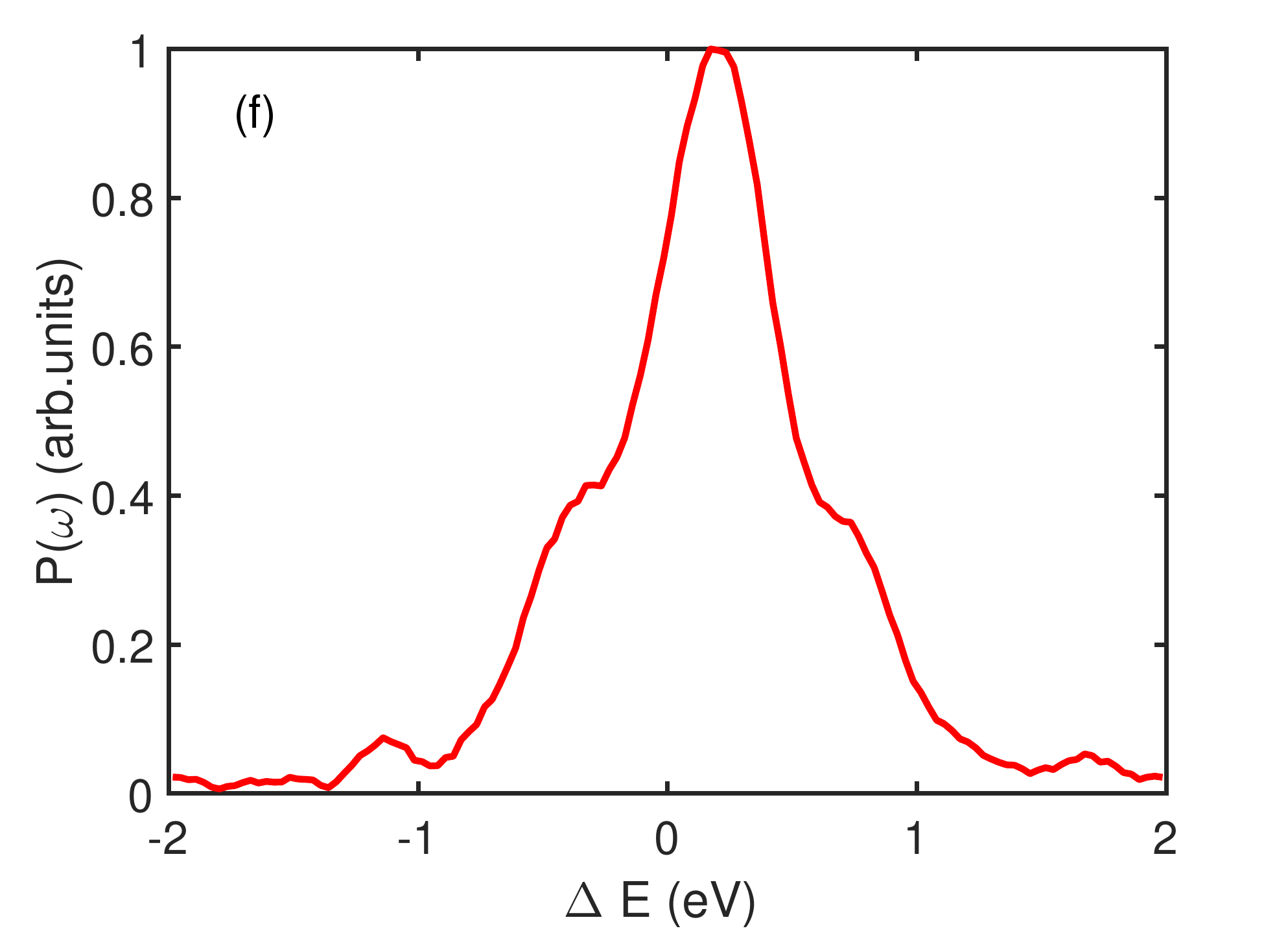}}
\caption{\label{fig_rb}The performance of real electron beam for XFELO with the 33\% crystal mirror output coupling, and for tapered FEL amplifier. The top row of panel shows (a) output pulse energy, (b) power profile (red solid line) and beam energy spread (green dashed line), (c) spectrum for XFELO, while the bottom row of panel (d), (e), (f) display the corresponding results for tapered FEL amplifier. }
\end{figure*}

As aforementioned, a chicane bypass is adjusted to provide proper delay to let the bunch head overlaps with the XFELO radiation and to wash out the XFELO-induced microbunching. Quadratic tapering model is chosen and carefully optimized to obtain maximum pulse energy for subsequent nearly 180 m length tapered FEL amplifier. The second row of the panel shows that the amplifier is able to generate 7.9 mJ pulse energy with 0.92 TW peak power, 586 meV (FWHM) spectral bandwidth and 14 fs (FWHM) pulse duration. This corresponds to a time-bandwidth product of 2, which is larger than the value of Fourier transform limit (0.44) for Gaussian pulse profile and is mainly due to the chirped electron beam we used. 

\section{\label{sec:level4}Start-to-end simulation of SCLF}
\begin{figure*} 
\centering
\subfigure{\includegraphics[width=5.5cm]{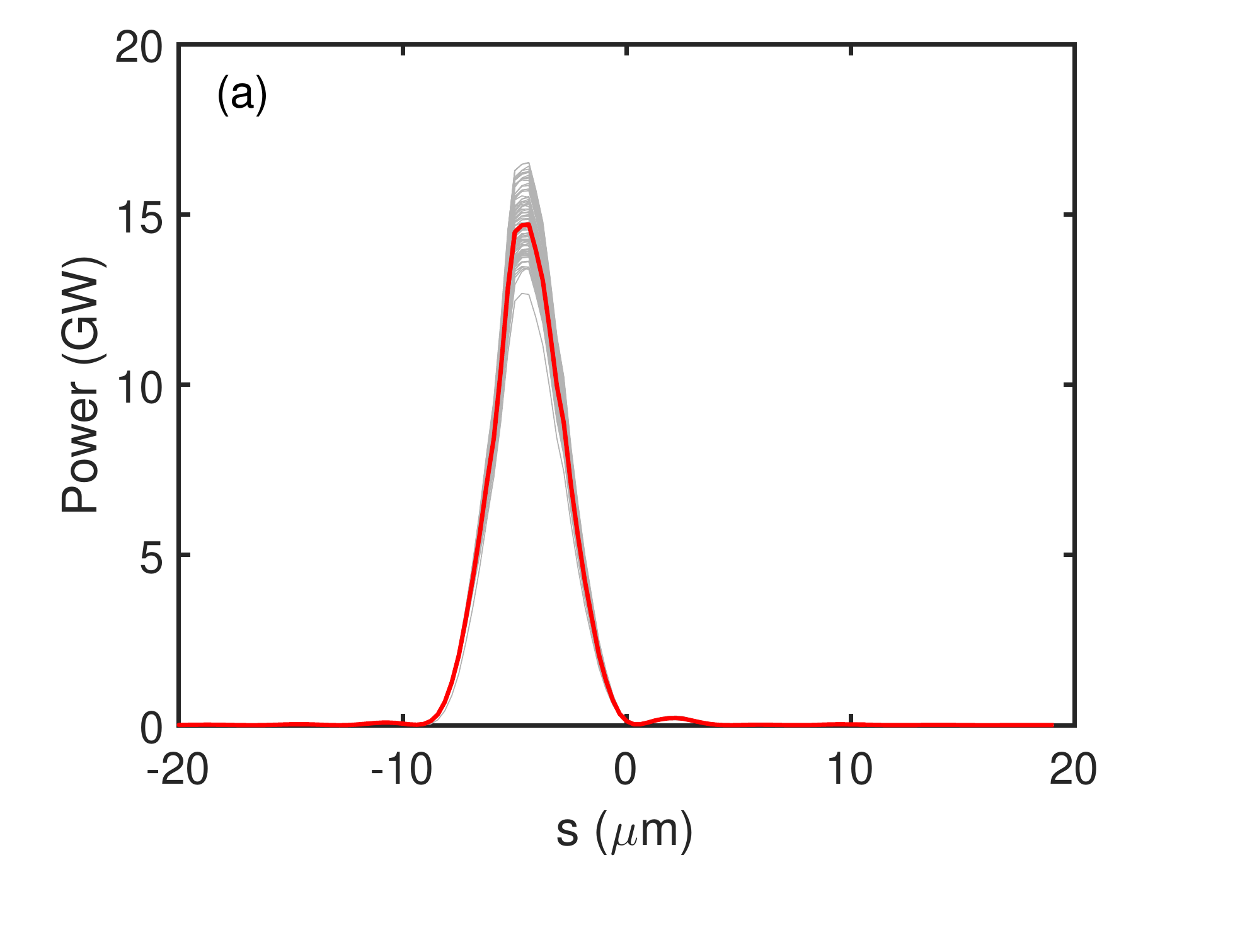}}
\subfigure{\includegraphics[width=5.5cm]{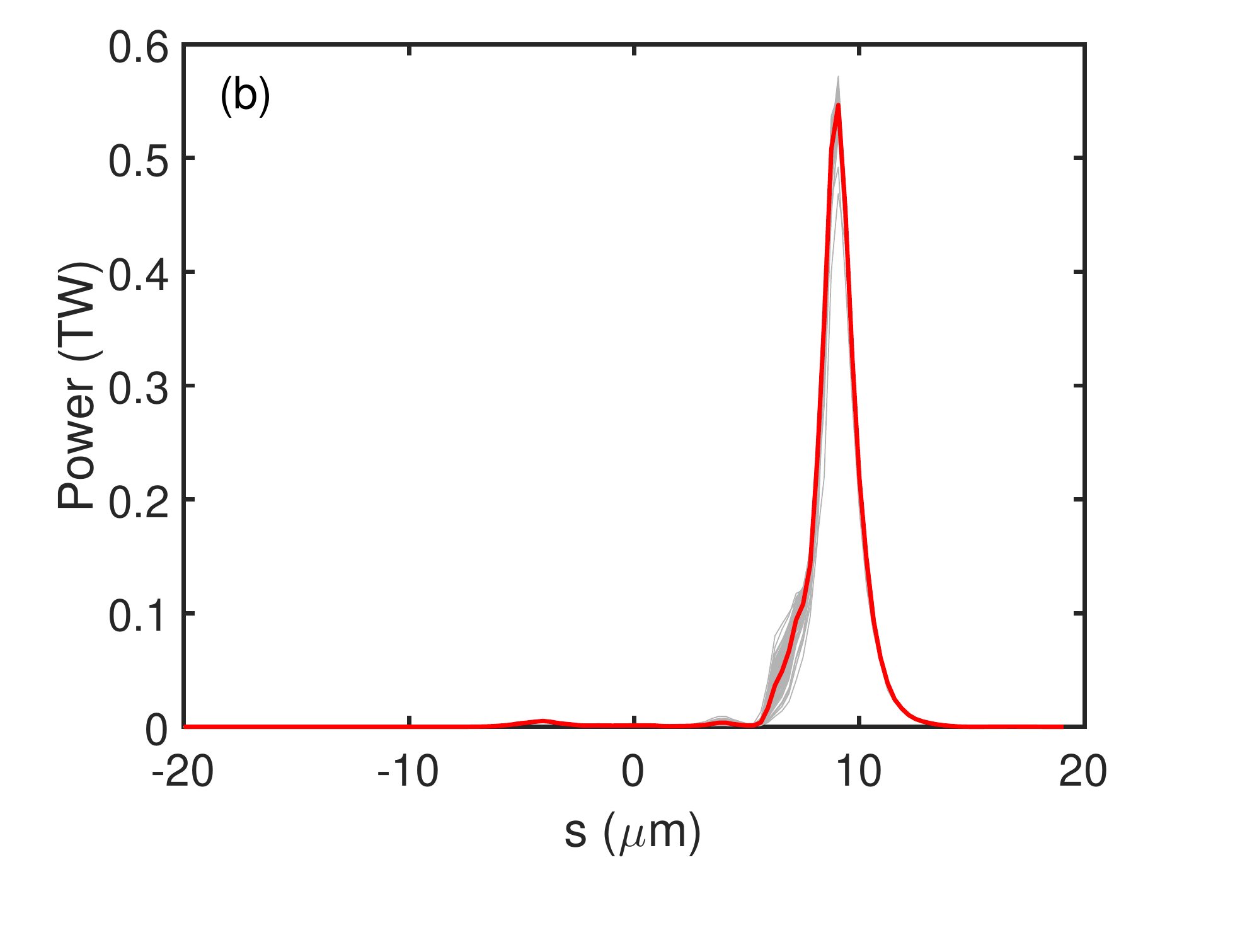}}
\subfigure{\includegraphics[width=5.5cm]{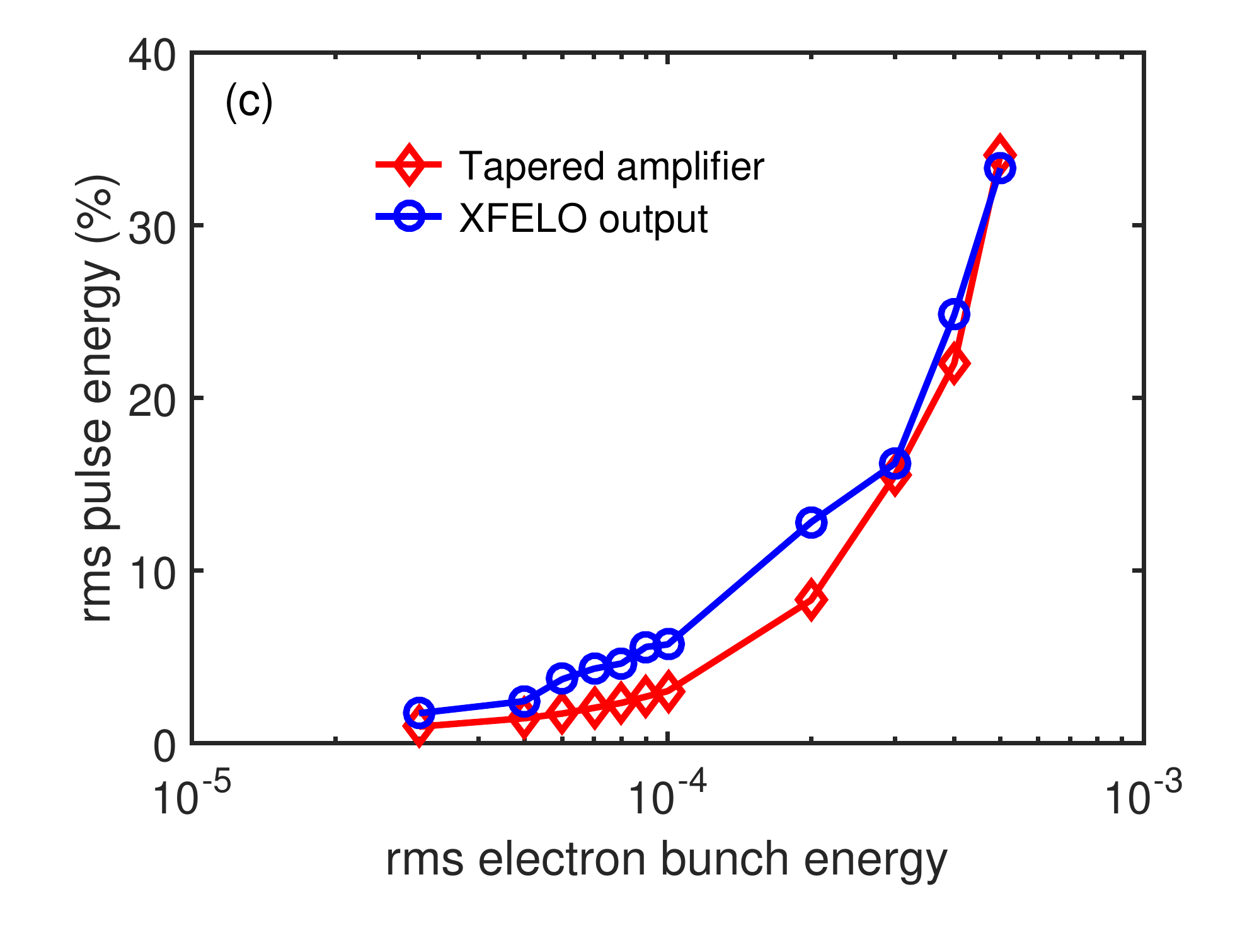}}
\caption{\label{stable}The output radiation of (a) XFELO and (b) tapered FEL amplifier. The gray lines refer to 100 runs with $10^{-4}$ rms beam energy jitter, and red line is the average value. Subfigure (c) shows the rms of output energy with different rms beam energy jitter.}
\end{figure*}
The performances of XFELO seeded FEL are investigated by start-to-end simulations of SCLF. The electron beam dynamics simulation in the photon-injector is carried out by ASTRA \cite{flottmann2011astra} with space charge effects taken into account. ELEGANT \cite{borland2000advanced} is then used for simulation in the reminder of Linac. For pursuing shorter gain length of SASE FEL, the electrons are accelerated off-crest in the superconducting Linac and possess large energy chirp, then the bunch is compressed longitudinally inside a chicane compressor to obtain high peak current. In order to achieve high efficient lasing of high-gain FELs, the residual energy chirp compensation is of paramount importance and several techniques are available for reaching this goal, i.e., off-crest dechirping and passively dechirping through Linac cavity wakefield. For SCLF using low-frequency superconducting technology, however, the wakefield is too weak to remove the residual chirp, and actively dechirping by running the beam off-crest in a Linac is an inefficient and costly option. Thus SCLF takes advantages of an inexpensive 12 m length corrugated pipe as a beam dechirper \cite{bane2012corrugated,antipov2012experimental,deng2014experimental}. 

In this paper, however, the introduced chirp is used for XFELO seeded FEL operation strictly without further chirp removal devices. The phase space electrons distribution (a) and the corresponding current profile (b) are presented in Fig.~\ref{fig_phase}.  For a given chirp $\alpha=1.3\times10^{-3}$, using the strategies mentioned above, the thickness of diamond crystal is adjusted to obtain 33\% output coupling for single segment undulator XFELO simulations. The natural ``double-horn'' of electron beam is proved to be quite suitable for our scheme, since one horn is lasing inside the XFELO while the other is well-preserved for lasing in FEL amplifier. The energy chirp required is for the head horn to be outside the gain bandwidth while the tail horn corresponding to maximum single pass gain, and the influence of small current between them could be ignored, thus leads to relative small optimized energy chirp. In addition, the large energy spread of the ``double horn'' current is not as harmful to XFELO seeded FEL as self-seeding scheme, since neither the XFELO, which works in the small signal regime, nor the tapered FEL amplifier is sensitive to electron beam energy spread. 

The simulation results of XFELO and tapered FEL amplifier for 5 keV photons are shown in the upper and bottom row of Fig.~\ref{fig_rb}, which demonstrate that the tail horn produces a nearly 190 $\mu$J, 14.5 GW XFELO output, with 0.3 eV (FWHM) spectral bandwidth. The energy spread corresponding to the bunch tail increases, while the bunch head remains unchanged. The XFELO output is amplified to 3.8 mJ, 0.55 TW, with 0.66 eV (FWHM) spectral bandwidth and 5.5 fs (FWHM) temporal duration. This corresponds to a time-bandwidth product of 0.88, which is twice the value of Fourier transform limit (0.44) for Gaussian pulse profile. Due to the relatively small bunch energy chirp as well as short radiation duration, the output pulse longitudinal coherence is better than the flat-top current case mentioned above. XFELO seeded FEL is able to generate 5 keV X-ray photons with $3.35\times10^{35}\,(photons/s/mm^2/mrad^2/0.1\% BW)$ peak brilliance, which is two order of magnitude larger than SASE FEL brilliance $1.1\times10^{33}\,(photons/s/mm^2/mrad^2/0.1\% BW)$ produced by the same electron and undulator parameters.

\section{\label{sec:level5}Output energy stability}
To illustrate the stability of XFELO seeded FEL in real machine, 100 start-to-end runs are carried out for each electron beam energy jitter rms value. The results of XFELO (a) and tapered amplifier (b) with rms equals to $10^{-4}$ are shown in Fig.~\ref{stable}. The gray lines refer to 100 runs and the red line is the average value of them. The results are summarized in Fig.~\ref{stable}(c), which indicates that with the normal $10^{-4}$ rms beam energy jitter in SCLF, XFELO seeded FEL is able to generate high brightness X-ray pulse with fluctuations as low as 3\%. Note that the final output is more resistant to bunch energy jitter than XFELO, further analysis reveals that this is due to the relatively broader gain bandwidth of tapered FEL amplifier in respect XFELO in our simulations.

\section{\label{sec:level6}Conclusion}
The chirped-beam XFELO seeded FEL is a novel strategy to generate high pulse energy, fully coherent X-ray beam. The XFELO output provides a fully coherent, high power, stable seed radiation, which means a favorable ``bucket'' to trap electrons in phase space and to maintain consistent energy extraction in tapered FEL amplifier. The chirped beam is optimized for ensuring the bunch tail lases effectively in XFELO while the bunch head is preserved for following tapered FEL amplifier. It is demonstrated that with optimized parameters, the new scheme is able to generate a few mJ, TW level fully coherent 5 keV X-ray radiation.

Taking advantages of the natural remnant energy chirp after chicane compressor, a start-to-end simulation is conducted using the basic parameters in SCLF. The ``double horn'' current is proved to be suitable for XFELO seeded FEL operation. Simulation results demonstrate that the new scheme is able to generate 3.8 mJ pulse energy, 0.55 TW peak power, 5 keV photon energy X-ray pulses whose peak brilliance is two order of magnitude larger than SASE FEL. In addition, XFELO seeded FEL output is quite stable, rms pulse energy fluctuation is as small as 3\% for a typical $10^{-4}$ rms electron beam energy jitter. The ideal X-ray light source is expected to be valuable for some special experiments, i.e., X-ray single-shot diffraction imaging and coherent scattering techniques, which require high spectral brightness. The following work would be to analyze the influence of bandwidth of seeding radiation on its performances as well as the gain bandwidth of tapered FEL amplifier.  
\begin{acknowledgments}
The author would like to thank B.~Liu, D.~Wang and Z.~Zhao for helpful discussions on SCLF project; T. Liu for enthusiastic discussions on FEL tapering physics. This work was partially supported by the National Natural Science Foundation of China (11775293) and Ten Thousand Talent Program.
\end{acknowledgments}

\nocite{*}

\bibliographystyle{apsrev4-1}
\bibliography{reference}

\begin{thebibliography}{44}%
\makeatletter
\providecommand \@ifxundefined [1]{%
 \@ifx{#1\undefined}
}%
\providecommand \@ifnum [1]{%
 \ifnum #1\expandafter \@firstoftwo
 \else \expandafter \@secondoftwo
 \fi
}%
\providecommand \@ifx [1]{%
 \ifx #1\expandafter \@firstoftwo
 \else \expandafter \@secondoftwo
 \fi
}%
\providecommand \natexlab [1]{#1}%
\providecommand \enquote  [1]{``#1''}%
\providecommand \bibnamefont  [1]{#1}%
\providecommand \bibfnamefont [1]{#1}%
\providecommand \citenamefont [1]{#1}%
\providecommand \href@noop [0]{\@secondoftwo}%
\providecommand \href [0]{\begingroup \@sanitize@url \@href}%
\providecommand \@href[1]{\@@startlink{#1}\@@href}%
\providecommand \@@href[1]{\endgroup#1\@@endlink}%
\providecommand \@sanitize@url [0]{\catcode `\\12\catcode `\$12\catcode
  `\&12\catcode `\#12\catcode `\^12\catcode `\_12\catcode `\%12\relax}%
\providecommand \@@startlink[1]{}%
\providecommand \@@endlink[0]{}%
\providecommand \url  [0]{\begingroup\@sanitize@url \@url }%
\providecommand \@url [1]{\endgroup\@href {#1}{\urlprefix }}%
\providecommand \urlprefix  [0]{URL }%
\providecommand \Eprint [0]{\href }%
\providecommand \doibase [0]{http://dx.doi.org/}%
\providecommand \selectlanguage [0]{\@gobble}%
\providecommand \bibinfo  [0]{\@secondoftwo}%
\providecommand \bibfield  [0]{\@secondoftwo}%
\providecommand \translation [1]{[#1]}%
\providecommand \BibitemOpen [0]{}%
\providecommand \bibitemStop [0]{}%
\providecommand \bibitemNoStop [0]{.\EOS\space}%
\providecommand \EOS [0]{\spacefactor3000\relax}%
\providecommand \BibitemShut  [1]{\csname bibitem#1\endcsname}%
\let\auto@bib@innerbib\@empty
\bibitem [{\citenamefont {Kondratenko}\ and\ \citenamefont
  {Saldin}(1980)}]{kondratenko1980generating}%
  \BibitemOpen
  \bibfield  {author} {\bibinfo {author} {\bibfnamefont {A.}~\bibnamefont
  {Kondratenko}}\ and\ \bibinfo {author} {\bibfnamefont {E.}~\bibnamefont
  {Saldin}},\ }\href@noop {} {\bibfield  {journal} {\bibinfo  {journal} {Part.
  Accel.}\ }\textbf {\bibinfo {volume} {10}},\ \bibinfo {pages} {207} (\bibinfo
  {year} {1980})}\BibitemShut {NoStop}%
\bibitem [{\citenamefont {Bonifacio}\ \emph {et~al.}(1984)\citenamefont
  {Bonifacio}, \citenamefont {Pellegrini},\ and\ \citenamefont
  {Narducci}}]{bonifacio1984collective}%
  \BibitemOpen
  \bibfield  {author} {\bibinfo {author} {\bibfnamefont {R.}~\bibnamefont
  {Bonifacio}}, \bibinfo {author} {\bibfnamefont {C.}~\bibnamefont
  {Pellegrini}}, \ and\ \bibinfo {author} {\bibfnamefont {L.}~\bibnamefont
  {Narducci}},\ }\href@noop {} {\bibfield  {journal} {\bibinfo  {journal}
  {Optics Communications}\ }\textbf {\bibinfo {volume} {50}},\ \bibinfo {pages}
  {373} (\bibinfo {year} {1984})}\BibitemShut {NoStop}%
\bibitem [{\citenamefont {Emma}\ \emph {et~al.}(2010)\citenamefont {Emma},
  \citenamefont {Akre}, \citenamefont {Arthur}, \citenamefont {Bionta},
  \citenamefont {Bostedt}, \citenamefont {Bozek}, \citenamefont {Brachmann},
  \citenamefont {Bucksbaum}, \citenamefont {Coffee}, \citenamefont {Decker}
  \emph {et~al.}}]{emma2010first}%
  \BibitemOpen
  \bibfield  {author} {\bibinfo {author} {\bibfnamefont {P.}~\bibnamefont
  {Emma}}, \bibinfo {author} {\bibfnamefont {R.}~\bibnamefont {Akre}}, \bibinfo
  {author} {\bibfnamefont {J.}~\bibnamefont {Arthur}}, \bibinfo {author}
  {\bibfnamefont {R.}~\bibnamefont {Bionta}}, \bibinfo {author} {\bibfnamefont
  {C.}~\bibnamefont {Bostedt}}, \bibinfo {author} {\bibfnamefont
  {J.}~\bibnamefont {Bozek}}, \bibinfo {author} {\bibfnamefont
  {A.}~\bibnamefont {Brachmann}}, \bibinfo {author} {\bibfnamefont
  {P.}~\bibnamefont {Bucksbaum}}, \bibinfo {author} {\bibfnamefont
  {R.}~\bibnamefont {Coffee}}, \bibinfo {author} {\bibfnamefont {F.-J.}\
  \bibnamefont {Decker}},  \emph {et~al.},\ }\href@noop {} {\bibfield
  {journal} {\bibinfo  {journal} {Nature Photonics}\ }\textbf {\bibinfo
  {volume} {4}},\ \bibinfo {pages} {641} (\bibinfo {year} {2010})}\BibitemShut
  {NoStop}%
\bibitem [{\citenamefont {Pile}(2011)}]{pile2011x}%
  \BibitemOpen
  \bibfield  {author} {\bibinfo {author} {\bibfnamefont {D.}~\bibnamefont
  {Pile}},\ }\href@noop {} {\bibfield  {journal} {\bibinfo  {journal} {Nature
  Photonics}\ }\textbf {\bibinfo {volume} {5}},\ \bibinfo {pages} {456}
  (\bibinfo {year} {2011})}\BibitemShut {NoStop}%
\bibitem [{\citenamefont {Kang}\ \emph {et~al.}(2013)\citenamefont {Kang},
  \citenamefont {Kim},\ and\ \citenamefont {Ko}}]{kang2013current}%
  \BibitemOpen
  \bibfield  {author} {\bibinfo {author} {\bibfnamefont {H.-S.}\ \bibnamefont
  {Kang}}, \bibinfo {author} {\bibfnamefont {K.-W.}\ \bibnamefont {Kim}}, \
  and\ \bibinfo {author} {\bibfnamefont {I.~S.}\ \bibnamefont {Ko}},\ }in\
  \href@noop {} {\emph {\bibinfo {booktitle} {Proceedings of 4th International
  Particle Accelerator Conference}}}\ (\bibinfo {organization} {JACOW,
  Shanghai, China},\ \bibinfo {year} {2013})\ p.\ \bibinfo {pages}
  {2074}\BibitemShut {NoStop}%
\bibitem [{\citenamefont {Altarelli}\ \emph {et~al.}(2006)\citenamefont
  {Altarelli}, \citenamefont {Brinkmann}, \citenamefont {Chergui},
  \citenamefont {Decking}, \citenamefont {Dobson}, \citenamefont
  {D{\"u}sterer}, \citenamefont {Gr{\"u}bel}, \citenamefont {Graeff},
  \citenamefont {Graafsma}, \citenamefont {Hajdu} \emph
  {et~al.}}]{altarelli2006european}%
  \BibitemOpen
  \bibfield  {author} {\bibinfo {author} {\bibfnamefont {M.}~\bibnamefont
  {Altarelli}}, \bibinfo {author} {\bibfnamefont {R.}~\bibnamefont
  {Brinkmann}}, \bibinfo {author} {\bibfnamefont {M.}~\bibnamefont {Chergui}},
  \bibinfo {author} {\bibfnamefont {W.}~\bibnamefont {Decking}}, \bibinfo
  {author} {\bibfnamefont {B.}~\bibnamefont {Dobson}}, \bibinfo {author}
  {\bibfnamefont {S.}~\bibnamefont {D{\"u}sterer}}, \bibinfo {author}
  {\bibfnamefont {G.}~\bibnamefont {Gr{\"u}bel}}, \bibinfo {author}
  {\bibfnamefont {W.}~\bibnamefont {Graeff}}, \bibinfo {author} {\bibfnamefont
  {H.}~\bibnamefont {Graafsma}}, \bibinfo {author} {\bibfnamefont
  {J.}~\bibnamefont {Hajdu}},  \emph {et~al.},\ }\href@noop {} {\bibfield
  {journal} {\bibinfo  {journal} {Technical Design Report, DESY}\ }\textbf
  {\bibinfo {volume} {97}},\ \bibinfo {pages} {1} (\bibinfo {year}
  {2006})}\BibitemShut {NoStop}%
\bibitem [{\citenamefont {Bostedt}\ \emph {et~al.}(2016)\citenamefont
  {Bostedt}, \citenamefont {Boutet}, \citenamefont {Fritz}, \citenamefont
  {Huang}, \citenamefont {Lee}, \citenamefont {Lemke}, \citenamefont {Robert},
  \citenamefont {Schlotter}, \citenamefont {Turner},\ and\ \citenamefont
  {Williams}}]{bostedt2016linac}%
  \BibitemOpen
  \bibfield  {author} {\bibinfo {author} {\bibfnamefont {C.}~\bibnamefont
  {Bostedt}}, \bibinfo {author} {\bibfnamefont {S.}~\bibnamefont {Boutet}},
  \bibinfo {author} {\bibfnamefont {D.~M.}\ \bibnamefont {Fritz}}, \bibinfo
  {author} {\bibfnamefont {Z.}~\bibnamefont {Huang}}, \bibinfo {author}
  {\bibfnamefont {H.~J.}\ \bibnamefont {Lee}}, \bibinfo {author} {\bibfnamefont
  {H.~T.}\ \bibnamefont {Lemke}}, \bibinfo {author} {\bibfnamefont
  {A.}~\bibnamefont {Robert}}, \bibinfo {author} {\bibfnamefont {W.~F.}\
  \bibnamefont {Schlotter}}, \bibinfo {author} {\bibfnamefont {J.~J.}\
  \bibnamefont {Turner}}, \ and\ \bibinfo {author} {\bibfnamefont {G.~J.}\
  \bibnamefont {Williams}},\ }\href@noop {} {\bibfield  {journal} {\bibinfo
  {journal} {Reviews of Modern Physics}\ }\textbf {\bibinfo {volume} {88}},\
  \bibinfo {pages} {015007} (\bibinfo {year} {2016})}\BibitemShut {NoStop}%
\bibitem [{\citenamefont {Chapman}\ \emph {et~al.}(2011)\citenamefont
  {Chapman}, \citenamefont {Fromme}, \citenamefont {Barty}, \citenamefont
  {White}, \citenamefont {Kirian}, \citenamefont {Aquila}, \citenamefont
  {Hunter}, \citenamefont {Schulz}, \citenamefont {DePonte}, \citenamefont
  {Weierstall} \emph {et~al.}}]{chapman2011femtosecond}%
  \BibitemOpen
  \bibfield  {author} {\bibinfo {author} {\bibfnamefont {H.~N.}\ \bibnamefont
  {Chapman}}, \bibinfo {author} {\bibfnamefont {P.}~\bibnamefont {Fromme}},
  \bibinfo {author} {\bibfnamefont {A.}~\bibnamefont {Barty}}, \bibinfo
  {author} {\bibfnamefont {T.~A.}\ \bibnamefont {White}}, \bibinfo {author}
  {\bibfnamefont {R.~A.}\ \bibnamefont {Kirian}}, \bibinfo {author}
  {\bibfnamefont {A.}~\bibnamefont {Aquila}}, \bibinfo {author} {\bibfnamefont
  {M.~S.}\ \bibnamefont {Hunter}}, \bibinfo {author} {\bibfnamefont
  {J.}~\bibnamefont {Schulz}}, \bibinfo {author} {\bibfnamefont {D.~P.}\
  \bibnamefont {DePonte}}, \bibinfo {author} {\bibfnamefont {U.}~\bibnamefont
  {Weierstall}},  \emph {et~al.},\ }\href@noop {} {\bibfield  {journal}
  {\bibinfo  {journal} {Nature}\ }\textbf {\bibinfo {volume} {470}},\ \bibinfo
  {pages} {73} (\bibinfo {year} {2011})}\BibitemShut {NoStop}%
\bibitem [{\citenamefont {Seibert}\ \emph {et~al.}(2011)\citenamefont
  {Seibert}, \citenamefont {Ekeberg}, \citenamefont {Maia}, \citenamefont
  {Svenda}, \citenamefont {Andreasson}, \citenamefont {J{\"o}nsson},
  \citenamefont {Odi{\'c}}, \citenamefont {Iwan}, \citenamefont {Rocker},
  \citenamefont {Westphal} \emph {et~al.}}]{seibert2011single}%
  \BibitemOpen
  \bibfield  {author} {\bibinfo {author} {\bibfnamefont {M.~M.}\ \bibnamefont
  {Seibert}}, \bibinfo {author} {\bibfnamefont {T.}~\bibnamefont {Ekeberg}},
  \bibinfo {author} {\bibfnamefont {F.~R.}\ \bibnamefont {Maia}}, \bibinfo
  {author} {\bibfnamefont {M.}~\bibnamefont {Svenda}}, \bibinfo {author}
  {\bibfnamefont {J.}~\bibnamefont {Andreasson}}, \bibinfo {author}
  {\bibfnamefont {O.}~\bibnamefont {J{\"o}nsson}}, \bibinfo {author}
  {\bibfnamefont {D.}~\bibnamefont {Odi{\'c}}}, \bibinfo {author}
  {\bibfnamefont {B.}~\bibnamefont {Iwan}}, \bibinfo {author} {\bibfnamefont
  {A.}~\bibnamefont {Rocker}}, \bibinfo {author} {\bibfnamefont
  {D.}~\bibnamefont {Westphal}},  \emph {et~al.},\ }\href@noop {} {\bibfield
  {journal} {\bibinfo  {journal} {Nature}\ }\textbf {\bibinfo {volume} {470}},\
  \bibinfo {pages} {78} (\bibinfo {year} {2011})}\BibitemShut {NoStop}%
\bibitem [{\citenamefont {Kroll}\ \emph {et~al.}(1981)\citenamefont {Kroll},
  \citenamefont {Morton},\ and\ \citenamefont {Rosenbluth}}]{kroll1981free}%
  \BibitemOpen
  \bibfield  {author} {\bibinfo {author} {\bibfnamefont {N.}~\bibnamefont
  {Kroll}}, \bibinfo {author} {\bibfnamefont {P.}~\bibnamefont {Morton}}, \
  and\ \bibinfo {author} {\bibfnamefont {M.}~\bibnamefont {Rosenbluth}},\
  }\href@noop {} {\bibfield  {journal} {\bibinfo  {journal} {IEEE Journal of
  Quantum Electronics}\ }\textbf {\bibinfo {volume} {17}},\ \bibinfo {pages}
  {1436} (\bibinfo {year} {1981})}\BibitemShut {NoStop}%
\bibitem [{\citenamefont {Schneidmiller}\ and\ \citenamefont
  {Yurkov}(2015)}]{schneidmiller2015optimization}%
  \BibitemOpen
  \bibfield  {author} {\bibinfo {author} {\bibfnamefont {E.~A.}\ \bibnamefont
  {Schneidmiller}}\ and\ \bibinfo {author} {\bibfnamefont {M.}~\bibnamefont
  {Yurkov}},\ }\href@noop {} {\bibfield  {journal} {\bibinfo  {journal}
  {Physical Review Special Topics-Accelerators and Beams}\ }\textbf {\bibinfo
  {volume} {18}},\ \bibinfo {pages} {030705} (\bibinfo {year}
  {2015})}\BibitemShut {NoStop}%
\bibitem [{\citenamefont {Jiao}\ \emph {et~al.}(2012)\citenamefont {Jiao},
  \citenamefont {Wu}, \citenamefont {Cai}, \citenamefont {Chao}, \citenamefont
  {Fawley}, \citenamefont {Frisch}, \citenamefont {Huang}, \citenamefont
  {Nuhn}, \citenamefont {Pellegrini},\ and\ \citenamefont
  {Reiche}}]{jiao2012modeling}%
  \BibitemOpen
  \bibfield  {author} {\bibinfo {author} {\bibfnamefont {Y.}~\bibnamefont
  {Jiao}}, \bibinfo {author} {\bibfnamefont {J.}~\bibnamefont {Wu}}, \bibinfo
  {author} {\bibfnamefont {Y.}~\bibnamefont {Cai}}, \bibinfo {author}
  {\bibfnamefont {A.}~\bibnamefont {Chao}}, \bibinfo {author} {\bibfnamefont
  {W.}~\bibnamefont {Fawley}}, \bibinfo {author} {\bibfnamefont
  {J.}~\bibnamefont {Frisch}}, \bibinfo {author} {\bibfnamefont
  {Z.}~\bibnamefont {Huang}}, \bibinfo {author} {\bibfnamefont {H.-D.}\
  \bibnamefont {Nuhn}}, \bibinfo {author} {\bibfnamefont {C.}~\bibnamefont
  {Pellegrini}}, \ and\ \bibinfo {author} {\bibfnamefont {S.}~\bibnamefont
  {Reiche}},\ }\href@noop {} {\bibfield  {journal} {\bibinfo  {journal}
  {Physical Review Special Topics-Accelerators and Beams}\ }\textbf {\bibinfo
  {volume} {15}},\ \bibinfo {pages} {050704} (\bibinfo {year}
  {2012})}\BibitemShut {NoStop}%
\bibitem [{\citenamefont {Orzechowski}\ \emph {et~al.}(1986)\citenamefont
  {Orzechowski}, \citenamefont {Anderson}, \citenamefont {Clark}, \citenamefont
  {Fawley}, \citenamefont {Paul}, \citenamefont {Prosnitz}, \citenamefont
  {Scharlemann}, \citenamefont {Yarema}, \citenamefont {Hopkins}, \citenamefont
  {Sessler} \emph {et~al.}}]{orzechowski1986high}%
  \BibitemOpen
  \bibfield  {author} {\bibinfo {author} {\bibfnamefont {T.~J.}\ \bibnamefont
  {Orzechowski}}, \bibinfo {author} {\bibfnamefont {B.}~\bibnamefont
  {Anderson}}, \bibinfo {author} {\bibfnamefont {J.~C.}\ \bibnamefont {Clark}},
  \bibinfo {author} {\bibfnamefont {W.~M.}\ \bibnamefont {Fawley}}, \bibinfo
  {author} {\bibfnamefont {A.~C.}\ \bibnamefont {Paul}}, \bibinfo {author}
  {\bibfnamefont {D.}~\bibnamefont {Prosnitz}}, \bibinfo {author}
  {\bibfnamefont {E.~T.}\ \bibnamefont {Scharlemann}}, \bibinfo {author}
  {\bibfnamefont {S.~M.}\ \bibnamefont {Yarema}}, \bibinfo {author}
  {\bibfnamefont {D.~B.}\ \bibnamefont {Hopkins}}, \bibinfo {author}
  {\bibfnamefont {A.~M.}\ \bibnamefont {Sessler}},  \emph {et~al.},\
  }\href@noop {} {\bibfield  {journal} {\bibinfo  {journal} {Physical Review
  Letters}\ }\textbf {\bibinfo {volume} {57}},\ \bibinfo {pages} {2172}
  (\bibinfo {year} {1986})}\BibitemShut {NoStop}%
\bibitem [{\citenamefont {Wang}\ \emph {et~al.}(2009)\citenamefont {Wang},
  \citenamefont {Freund}, \citenamefont {Harder}, \citenamefont {Miner~Jr},
  \citenamefont {Murphy}, \citenamefont {Qian}, \citenamefont {Shen},\ and\
  \citenamefont {Yang}}]{wang2009efficiency}%
  \BibitemOpen
  \bibfield  {author} {\bibinfo {author} {\bibfnamefont {X.}~\bibnamefont
  {Wang}}, \bibinfo {author} {\bibfnamefont {H.}~\bibnamefont {Freund}},
  \bibinfo {author} {\bibfnamefont {D.}~\bibnamefont {Harder}}, \bibinfo
  {author} {\bibfnamefont {W.}~\bibnamefont {Miner~Jr}}, \bibinfo {author}
  {\bibfnamefont {J.}~\bibnamefont {Murphy}}, \bibinfo {author} {\bibfnamefont
  {H.}~\bibnamefont {Qian}}, \bibinfo {author} {\bibfnamefont {Y.}~\bibnamefont
  {Shen}}, \ and\ \bibinfo {author} {\bibfnamefont {X.}~\bibnamefont {Yang}},\
  }\href@noop {} {\bibfield  {journal} {\bibinfo  {journal} {Physical Review
  Letters}\ }\textbf {\bibinfo {volume} {103}},\ \bibinfo {pages} {154801}
  (\bibinfo {year} {2009})}\BibitemShut {NoStop}%
\bibitem [{\citenamefont {Huang}\ and\ \citenamefont
  {Stupakov}(2005)}]{huang2005free}%
  \BibitemOpen
  \bibfield  {author} {\bibinfo {author} {\bibfnamefont {Z.}~\bibnamefont
  {Huang}}\ and\ \bibinfo {author} {\bibfnamefont {G.}~\bibnamefont
  {Stupakov}},\ }\href@noop {} {\bibfield  {journal} {\bibinfo  {journal}
  {Physical Review Special Topics-Accelerators and Beams}\ }\textbf {\bibinfo
  {volume} {8}},\ \bibinfo {pages} {040702} (\bibinfo {year}
  {2005})}\BibitemShut {NoStop}%
\bibitem [{\citenamefont {Saldin}\ \emph {et~al.}(2006)\citenamefont {Saldin},
  \citenamefont {Schneidmiller},\ and\ \citenamefont
  {Yurkov}}]{saldin2006self}%
  \BibitemOpen
  \bibfield  {author} {\bibinfo {author} {\bibfnamefont {E.~L.}\ \bibnamefont
  {Saldin}}, \bibinfo {author} {\bibfnamefont {E.~A.}\ \bibnamefont
  {Schneidmiller}}, \ and\ \bibinfo {author} {\bibfnamefont {M.~V.}\
  \bibnamefont {Yurkov}},\ }\href@noop {} {\bibfield  {journal} {\bibinfo
  {journal} {Physical Review Special Topics-Accelerators and Beams}\ }\textbf
  {\bibinfo {volume} {9}},\ \bibinfo {pages} {050702} (\bibinfo {year}
  {2006})}\BibitemShut {NoStop}%
\bibitem [{\citenamefont {Geloni}\ \emph {et~al.}(2011)\citenamefont {Geloni},
  \citenamefont {Kocharyan},\ and\ \citenamefont {Saldin}}]{geloni2011novel}%
  \BibitemOpen
  \bibfield  {author} {\bibinfo {author} {\bibfnamefont {G.}~\bibnamefont
  {Geloni}}, \bibinfo {author} {\bibfnamefont {V.}~\bibnamefont {Kocharyan}}, \
  and\ \bibinfo {author} {\bibfnamefont {E.}~\bibnamefont {Saldin}},\
  }\href@noop {} {\bibfield  {journal} {\bibinfo  {journal} {Journal of Modern
  Optics}\ }\textbf {\bibinfo {volume} {58}},\ \bibinfo {pages} {1391}
  (\bibinfo {year} {2011})}\BibitemShut {NoStop}%
\bibitem [{\citenamefont {Amann}\ \emph {et~al.}(2012)\citenamefont {Amann},
  \citenamefont {Berg}, \citenamefont {Blank}, \citenamefont {Decker},
  \citenamefont {Ding}, \citenamefont {Emma}, \citenamefont {Feng},
  \citenamefont {Frisch}, \citenamefont {Fritz}, \citenamefont {Hastings} \emph
  {et~al.}}]{amann2012demonstration}%
  \BibitemOpen
  \bibfield  {author} {\bibinfo {author} {\bibfnamefont {J.}~\bibnamefont
  {Amann}}, \bibinfo {author} {\bibfnamefont {W.}~\bibnamefont {Berg}},
  \bibinfo {author} {\bibfnamefont {V.}~\bibnamefont {Blank}}, \bibinfo
  {author} {\bibfnamefont {F.-J.}\ \bibnamefont {Decker}}, \bibinfo {author}
  {\bibfnamefont {Y.}~\bibnamefont {Ding}}, \bibinfo {author} {\bibfnamefont
  {P.}~\bibnamefont {Emma}}, \bibinfo {author} {\bibfnamefont {Y.}~\bibnamefont
  {Feng}}, \bibinfo {author} {\bibfnamefont {J.}~\bibnamefont {Frisch}},
  \bibinfo {author} {\bibfnamefont {D.}~\bibnamefont {Fritz}}, \bibinfo
  {author} {\bibfnamefont {J.}~\bibnamefont {Hastings}},  \emph {et~al.},\
  }\href@noop {} {\bibfield  {journal} {\bibinfo  {journal} {Nature Photonics}\
  }\textbf {\bibinfo {volume} {6}},\ \bibinfo {pages} {693} (\bibinfo {year}
  {2012})}\BibitemShut {NoStop}%
\bibitem [{\citenamefont {Xiang}\ \emph {et~al.}(2013)\citenamefont {Xiang},
  \citenamefont {Ding}, \citenamefont {Huang},\ and\ \citenamefont
  {Deng}}]{xiang2013purified}%
  \BibitemOpen
  \bibfield  {author} {\bibinfo {author} {\bibfnamefont {D.}~\bibnamefont
  {Xiang}}, \bibinfo {author} {\bibfnamefont {Y.}~\bibnamefont {Ding}},
  \bibinfo {author} {\bibfnamefont {Z.}~\bibnamefont {Huang}}, \ and\ \bibinfo
  {author} {\bibfnamefont {H.}~\bibnamefont {Deng}},\ }\href@noop {} {\bibfield
   {journal} {\bibinfo  {journal} {Physical Review Special Topics-Accelerators
  and Beams}\ }\textbf {\bibinfo {volume} {16}},\ \bibinfo {pages} {010703}
  (\bibinfo {year} {2013})}\BibitemShut {NoStop}%
\bibitem [{\citenamefont {McNeil}\ \emph {et~al.}(2013)\citenamefont {McNeil},
  \citenamefont {Thompson},\ and\ \citenamefont
  {Dunning}}]{mcneil2013transform}%
  \BibitemOpen
  \bibfield  {author} {\bibinfo {author} {\bibfnamefont {B.}~\bibnamefont
  {McNeil}}, \bibinfo {author} {\bibfnamefont {N.}~\bibnamefont {Thompson}}, \
  and\ \bibinfo {author} {\bibfnamefont {D.}~\bibnamefont {Dunning}},\
  }\href@noop {} {\bibfield  {journal} {\bibinfo  {journal} {Physical Review
  Letters}\ }\textbf {\bibinfo {volume} {110}},\ \bibinfo {pages} {134802}
  (\bibinfo {year} {2013})}\BibitemShut {NoStop}%
\bibitem [{\citenamefont {Wu}\ \emph {et~al.}(2012)\citenamefont {Wu},
  \citenamefont {Marinelli},\ and\ \citenamefont
  {Pellegrini}}]{wu2012generation}%
  \BibitemOpen
  \bibfield  {author} {\bibinfo {author} {\bibfnamefont {J.}~\bibnamefont
  {Wu}}, \bibinfo {author} {\bibfnamefont {A.}~\bibnamefont {Marinelli}}, \
  and\ \bibinfo {author} {\bibfnamefont {C.}~\bibnamefont {Pellegrini}},\ }in\
  \href@noop {} {\emph {\bibinfo {booktitle} {Proceedings of the 34th
  International Free Electron Laser Conference}}}\ (\bibinfo {organization}
  {JACOW, Nara, Japan},\ \bibinfo {year} {2012})\ p.\ \bibinfo {pages}
  {237}\BibitemShut {NoStop}%
\bibitem [{\citenamefont {Yu}(1991)}]{yu1991generation}%
  \BibitemOpen
  \bibfield  {author} {\bibinfo {author} {\bibfnamefont {L.~H.}\ \bibnamefont
  {Yu}},\ }\href@noop {} {\bibfield  {journal} {\bibinfo  {journal} {Physical
  Review A}\ }\textbf {\bibinfo {volume} {44}},\ \bibinfo {pages} {5178}
  (\bibinfo {year} {1991})}\BibitemShut {NoStop}%
\bibitem [{\citenamefont {Stupakov}(2009)}]{stupakov2009using}%
  \BibitemOpen
  \bibfield  {author} {\bibinfo {author} {\bibfnamefont {G.}~\bibnamefont
  {Stupakov}},\ }\href@noop {} {\bibfield  {journal} {\bibinfo  {journal}
  {Physical Review Letters}\ }\textbf {\bibinfo {volume} {102}},\ \bibinfo
  {pages} {074801} (\bibinfo {year} {2009})}\BibitemShut {NoStop}%
\bibitem [{\citenamefont {Deng}\ and\ \citenamefont
  {Feng}(2013{\natexlab{a}})}]{deng2013using}%
  \BibitemOpen
  \bibfield  {author} {\bibinfo {author} {\bibfnamefont {H.}~\bibnamefont
  {Deng}}\ and\ \bibinfo {author} {\bibfnamefont {C.}~\bibnamefont {Feng}},\
  }\href@noop {} {\bibfield  {journal} {\bibinfo  {journal} {Physical Review
  Letters}\ }\textbf {\bibinfo {volume} {111}},\ \bibinfo {pages} {084801}
  (\bibinfo {year} {2013}{\natexlab{a}})}\BibitemShut {NoStop}%
\bibitem [{\citenamefont {Deng}\ and\ \citenamefont
  {Feng}(2013{\natexlab{b}})}]{denghigh}%
  \BibitemOpen
  \bibfield  {author} {\bibinfo {author} {\bibfnamefont {H.}~\bibnamefont
  {Deng}}\ and\ \bibinfo {author} {\bibfnamefont {C.}~\bibnamefont {Feng}},\
  }in\ \href@noop {} {\emph {\bibinfo {booktitle} {Proceedings of the 4th
  International Particle Accelerator Conference}}}\ (\bibinfo {address} {JACOW,
  Shanghai, China},\ \bibinfo {year} {2013})\ p.\ \bibinfo {pages}
  {1214}\BibitemShut {NoStop}%
\bibitem [{\citenamefont {Kim}\ \emph {et~al.}(2008)\citenamefont {Kim},
  \citenamefont {Shvyd’ko},\ and\ \citenamefont {Reiche}}]{kim2008proposal}%
  \BibitemOpen
  \bibfield  {author} {\bibinfo {author} {\bibfnamefont {K.-J.}\ \bibnamefont
  {Kim}}, \bibinfo {author} {\bibfnamefont {Y.}~\bibnamefont {Shvyd’ko}}, \
  and\ \bibinfo {author} {\bibfnamefont {S.}~\bibnamefont {Reiche}},\
  }\href@noop {} {\bibfield  {journal} {\bibinfo  {journal} {Physical Review
  Letters}\ }\textbf {\bibinfo {volume} {100}},\ \bibinfo {pages} {244802}
  (\bibinfo {year} {2008})}\BibitemShut {NoStop}%
\bibitem [{\citenamefont {Dai}\ \emph {et~al.}(2012)\citenamefont {Dai},
  \citenamefont {Deng},\ and\ \citenamefont {Dai}}]{dai2012proposal}%
  \BibitemOpen
  \bibfield  {author} {\bibinfo {author} {\bibfnamefont {J.}~\bibnamefont
  {Dai}}, \bibinfo {author} {\bibfnamefont {H.}~\bibnamefont {Deng}}, \ and\
  \bibinfo {author} {\bibfnamefont {Z.}~\bibnamefont {Dai}},\ }\href@noop {}
  {\bibfield  {journal} {\bibinfo  {journal} {Physical Review Letters}\
  }\textbf {\bibinfo {volume} {108}},\ \bibinfo {pages} {034802} (\bibinfo
  {year} {2012})}\BibitemShut {NoStop}%
\bibitem [{\citenamefont {Li}\ and\ \citenamefont
  {Deng}(2017)}]{kai2017systematical}%
  \BibitemOpen
  \bibfield  {author} {\bibinfo {author} {\bibfnamefont {K.}~\bibnamefont
  {Li}}\ and\ \bibinfo {author} {\bibfnamefont {H.}~\bibnamefont {Deng}},\
  }\href@noop {} {\bibfield  {journal} {\bibinfo  {journal} {arXiv preprint
  arXiv:1706.06338}\ } (\bibinfo {year} {2017})}\BibitemShut {NoStop}%
\bibitem [{\citenamefont {Shvyd'Ko}\ \emph {et~al.}(2010)\citenamefont
  {Shvyd'Ko}, \citenamefont {Stoupin}, \citenamefont {Cunsolo}, \citenamefont
  {Said},\ and\ \citenamefont {Huang}}]{shvyd2010high}%
  \BibitemOpen
  \bibfield  {author} {\bibinfo {author} {\bibfnamefont {Y.~V.}\ \bibnamefont
  {Shvyd'Ko}}, \bibinfo {author} {\bibfnamefont {S.}~\bibnamefont {Stoupin}},
  \bibinfo {author} {\bibfnamefont {A.}~\bibnamefont {Cunsolo}}, \bibinfo
  {author} {\bibfnamefont {A.~H.}\ \bibnamefont {Said}}, \ and\ \bibinfo
  {author} {\bibfnamefont {X.}~\bibnamefont {Huang}},\ }\href@noop {}
  {\bibfield  {journal} {\bibinfo  {journal} {Nature Physics}\ }\textbf
  {\bibinfo {volume} {6}},\ \bibinfo {pages} {196} (\bibinfo {year}
  {2010})}\BibitemShut {NoStop}%
\bibitem [{\citenamefont {Li}\ \emph {et~al.}(2017)\citenamefont {Li},
  \citenamefont {Song},\ and\ \citenamefont {Deng}}]{li2017simplified}%
  \BibitemOpen
  \bibfield  {author} {\bibinfo {author} {\bibfnamefont {K.}~\bibnamefont
  {Li}}, \bibinfo {author} {\bibfnamefont {M.}~\bibnamefont {Song}}, \ and\
  \bibinfo {author} {\bibfnamefont {H.}~\bibnamefont {Deng}},\ }\href@noop {}
  {\bibfield  {journal} {\bibinfo  {journal} {Physical Review Special
  Topics-Accelerators and Beams}\ }\textbf {\bibinfo {volume} {20}},\ \bibinfo
  {pages} {030702} (\bibinfo {year} {2017})}\BibitemShut {NoStop}%
\bibitem [{\citenamefont {Lindberg}\ \emph {et~al.}(2011)\citenamefont
  {Lindberg}, \citenamefont {Kim}, \citenamefont {Shvyd’ko},\ and\
  \citenamefont {Fawley}}]{lindberg2011performance}%
  \BibitemOpen
  \bibfield  {author} {\bibinfo {author} {\bibfnamefont {R.}~\bibnamefont
  {Lindberg}}, \bibinfo {author} {\bibfnamefont {K.-J.}\ \bibnamefont {Kim}},
  \bibinfo {author} {\bibfnamefont {Y.}~\bibnamefont {Shvyd’ko}}, \ and\
  \bibinfo {author} {\bibfnamefont {W.}~\bibnamefont {Fawley}},\ }\href@noop {}
  {\bibfield  {journal} {\bibinfo  {journal} {Physical Review Special
  Topics-Accelerators and Beams}\ }\textbf {\bibinfo {volume} {14}},\ \bibinfo
  {pages} {010701} (\bibinfo {year} {2011})}\BibitemShut {NoStop}%
\bibitem [{\citenamefont {Borland}\ \emph {et~al.}(2007)\citenamefont
  {Borland}, \citenamefont {Decker}, \citenamefont {Nassiri},\ and\
  \citenamefont {White}}]{borland2007configuration}%
  \BibitemOpen
  \bibfield  {author} {\bibinfo {author} {\bibfnamefont {M.}~\bibnamefont
  {Borland}}, \bibinfo {author} {\bibfnamefont {G.}~\bibnamefont {Decker}},
  \bibinfo {author} {\bibfnamefont {A.}~\bibnamefont {Nassiri}}, \ and\
  \bibinfo {author} {\bibfnamefont {A.}~\bibnamefont {White}},\ }in\ \href@noop
  {} {\emph {\bibinfo {booktitle} {Proceedings of Particle Accelerator
  Conference 07}}}\ (\bibinfo {organization} {IEEE, Albuquerque, NM USA},\
  \bibinfo {year} {2007})\ p.\ \bibinfo {pages} {1121}\BibitemShut {NoStop}%
\bibitem [{\citenamefont {Shvyd’ko}\ and\ \citenamefont
  {Lindberg}(2012)}]{shvyd2012spatiotemporal}%
  \BibitemOpen
  \bibfield  {author} {\bibinfo {author} {\bibfnamefont {Y.}~\bibnamefont
  {Shvyd’ko}}\ and\ \bibinfo {author} {\bibfnamefont {R.}~\bibnamefont
  {Lindberg}},\ }\href@noop {} {\bibfield  {journal} {\bibinfo  {journal}
  {Physical Review Special Topics-Accelerators and Beams}\ }\textbf {\bibinfo
  {volume} {15}},\ \bibinfo {pages} {100702} (\bibinfo {year}
  {2012})}\BibitemShut {NoStop}%
\bibitem [{\citenamefont {Pierini}\ \emph {et~al.}(2017)\citenamefont
  {Pierini}, \citenamefont {Bertucci}, \citenamefont {Bosotti}, \citenamefont
  {Chen}, \citenamefont {Maiano}, \citenamefont {Michelato}, \citenamefont
  {Monaco}, \citenamefont {Moretti}, \citenamefont {Pagani}, \citenamefont
  {Paparella} \emph {et~al.}}]{pierini2017fabrication}%
  \BibitemOpen
  \bibfield  {author} {\bibinfo {author} {\bibfnamefont {P.}~\bibnamefont
  {Pierini}}, \bibinfo {author} {\bibfnamefont {M.}~\bibnamefont {Bertucci}},
  \bibinfo {author} {\bibfnamefont {A.}~\bibnamefont {Bosotti}}, \bibinfo
  {author} {\bibfnamefont {J.}~\bibnamefont {Chen}}, \bibinfo {author}
  {\bibfnamefont {C.}~\bibnamefont {Maiano}}, \bibinfo {author} {\bibfnamefont
  {P.}~\bibnamefont {Michelato}}, \bibinfo {author} {\bibfnamefont
  {L.}~\bibnamefont {Monaco}}, \bibinfo {author} {\bibfnamefont
  {M.}~\bibnamefont {Moretti}}, \bibinfo {author} {\bibfnamefont
  {C.}~\bibnamefont {Pagani}}, \bibinfo {author} {\bibfnamefont
  {R.}~\bibnamefont {Paparella}},  \emph {et~al.},\ }\href@noop {} {\bibfield
  {journal} {\bibinfo  {journal} {Physical Review Special Topics-Accelerators
  and Beams}\ }\textbf {\bibinfo {volume} {20}},\ \bibinfo {pages} {042006}
  (\bibinfo {year} {2017})}\BibitemShut {NoStop}%
\bibitem [{\citenamefont {Burrill}\ \emph {et~al.}(2017)\citenamefont
  {Burrill}, \citenamefont {Davis}, \citenamefont {Gonnella}, \citenamefont
  {Grassellino}, \citenamefont {Melnychuk}, \citenamefont {Palczewski},
  \citenamefont {Ross},\ and\ \citenamefont {Zhao}}]{burrill2017vertical}%
  \BibitemOpen
  \bibfield  {author} {\bibinfo {author} {\bibfnamefont {A.}~\bibnamefont
  {Burrill}}, \bibinfo {author} {\bibfnamefont {K.}~\bibnamefont {Davis}},
  \bibinfo {author} {\bibfnamefont {D.}~\bibnamefont {Gonnella}}, \bibinfo
  {author} {\bibfnamefont {A.}~\bibnamefont {Grassellino}}, \bibinfo {author}
  {\bibfnamefont {O.}~\bibnamefont {Melnychuk}}, \bibinfo {author}
  {\bibfnamefont {A.}~\bibnamefont {Palczewski}}, \bibinfo {author}
  {\bibfnamefont {M.}~\bibnamefont {Ross}}, \ and\ \bibinfo {author}
  {\bibfnamefont {L.}~\bibnamefont {Zhao}},\ }in\ \href@noop {} {\emph
  {\bibinfo {booktitle} {Proceedings of 8th International Particle Accelerator
  Conference}}}\ (\bibinfo {organization} {JACOW, Copenhagen, Denmark},\
  \bibinfo {year} {2017})\ p.\ \bibinfo {pages} {1152}\BibitemShut {NoStop}%
\bibitem [{\citenamefont {Qin}\ \emph {et~al.}(2017)\citenamefont {Qin},
  \citenamefont {Kim}, \citenamefont {Lindberg},\ and\ \citenamefont
  {Wu}}]{weilun2017xfelo}%
  \BibitemOpen
  \bibfield  {author} {\bibinfo {author} {\bibfnamefont {W.}~\bibnamefont
  {Qin}}, \bibinfo {author} {\bibfnamefont {K.-J.}\ \bibnamefont {Kim}},
  \bibinfo {author} {\bibfnamefont {R.}~\bibnamefont {Lindberg}}, \ and\
  \bibinfo {author} {\bibfnamefont {J.}~\bibnamefont {Wu}},\ }in\ \href@noop {}
  {\emph {\bibinfo {booktitle} {Proceedings of 38th International Free Electron
  Laser Conference}}}\ (\bibinfo {organization} {JACOW, Santa Fe, NM USA},\
  \bibinfo {year} {2017})\BibitemShut {NoStop}%
\bibitem [{\citenamefont {Lutman}\ \emph {et~al.}(2016)\citenamefont {Lutman},
  \citenamefont {Maxwell}, \citenamefont {MacArthur}, \citenamefont {Guetg},
  \citenamefont {Berrah}, \citenamefont {Coffee}, \citenamefont {Ding},
  \citenamefont {Huang}, \citenamefont {Marinelli}, \citenamefont {Moeller}
  \emph {et~al.}}]{lutman2016fresh}%
  \BibitemOpen
  \bibfield  {author} {\bibinfo {author} {\bibfnamefont {A.~A.}\ \bibnamefont
  {Lutman}}, \bibinfo {author} {\bibfnamefont {T.~J.}\ \bibnamefont {Maxwell}},
  \bibinfo {author} {\bibfnamefont {J.~P.}\ \bibnamefont {MacArthur}}, \bibinfo
  {author} {\bibfnamefont {M.~W.}\ \bibnamefont {Guetg}}, \bibinfo {author}
  {\bibfnamefont {N.}~\bibnamefont {Berrah}}, \bibinfo {author} {\bibfnamefont
  {R.~N.}\ \bibnamefont {Coffee}}, \bibinfo {author} {\bibfnamefont
  {Y.}~\bibnamefont {Ding}}, \bibinfo {author} {\bibfnamefont {Z.}~\bibnamefont
  {Huang}}, \bibinfo {author} {\bibfnamefont {A.}~\bibnamefont {Marinelli}},
  \bibinfo {author} {\bibfnamefont {S.}~\bibnamefont {Moeller}},  \emph
  {et~al.},\ }\href@noop {} {\bibfield  {journal} {\bibinfo  {journal} {Nature
  Photonics}\ }\textbf {\bibinfo {volume} {10}},\ \bibinfo {pages} {745}
  (\bibinfo {year} {2016})}\BibitemShut {NoStop}%
\bibitem [{\citenamefont {Reiche}(1999)}]{reiche1999genesis}%
  \BibitemOpen
  \bibfield  {author} {\bibinfo {author} {\bibfnamefont {S.}~\bibnamefont
  {Reiche}},\ }\href@noop {} {\bibfield  {journal} {\bibinfo  {journal}
  {Nuclear Instruments and Methods in Physics Research Section A: Accelerators,
  Spectrometers, Detectors and Associated Equipment}\ }\textbf {\bibinfo
  {volume} {429}},\ \bibinfo {pages} {243} (\bibinfo {year}
  {1999})}\BibitemShut {NoStop}%
\bibitem [{\citenamefont {van~der Slot}\ \emph {et~al.}(2009)\citenamefont
  {van~der Slot}, \citenamefont {Freund}, \citenamefont {Miner~Jr},
  \citenamefont {Benson}, \citenamefont {Shinn},\ and\ \citenamefont
  {Boller}}]{van2009time}%
  \BibitemOpen
  \bibfield  {author} {\bibinfo {author} {\bibfnamefont {P.~J.}\ \bibnamefont
  {van~der Slot}}, \bibinfo {author} {\bibfnamefont {H.}~\bibnamefont
  {Freund}}, \bibinfo {author} {\bibfnamefont {W.}~\bibnamefont {Miner~Jr}},
  \bibinfo {author} {\bibfnamefont {S.}~\bibnamefont {Benson}}, \bibinfo
  {author} {\bibfnamefont {M.}~\bibnamefont {Shinn}}, \ and\ \bibinfo {author}
  {\bibfnamefont {K.-J.}\ \bibnamefont {Boller}},\ }\href@noop {} {\bibfield
  {journal} {\bibinfo  {journal} {Physical Review Letters}\ }\textbf {\bibinfo
  {volume} {102}},\ \bibinfo {pages} {244802} (\bibinfo {year}
  {2009})}\BibitemShut {NoStop}%
\bibitem [{\citenamefont {Fl{\"o}ttmann}\ \emph {et~al.}(2011)\citenamefont
  {Fl{\"o}ttmann} \emph {et~al.}}]{flottmann2011astra}%
  \BibitemOpen
  \bibfield  {author} {\bibinfo {author} {\bibfnamefont {K.}~\bibnamefont
  {Fl{\"o}ttmann}} \emph {et~al.},\ }\href@noop {} {\bibfield  {journal}
  {\bibinfo  {journal} {Manual, Version}\ }\textbf {\bibinfo {volume} {3}},\
  \bibinfo {pages} {2014} (\bibinfo {year} {2011})}\BibitemShut {NoStop}%
\bibitem [{\citenamefont {Borland}(2000)}]{borland2000advanced}%
  \BibitemOpen
  \bibfield  {author} {\bibinfo {author} {\bibfnamefont {M.}~\bibnamefont
  {Borland}},\ }\href@noop {} {\bibfield  {journal} {\bibinfo  {journal}
  {LS-287}\ } (\bibinfo {year} {2000})}\BibitemShut {NoStop}%
\bibitem [{\citenamefont {Bane}\ and\ \citenamefont
  {Stupakov}(2012)}]{bane2012corrugated}%
  \BibitemOpen
  \bibfield  {author} {\bibinfo {author} {\bibfnamefont {K.}~\bibnamefont
  {Bane}}\ and\ \bibinfo {author} {\bibfnamefont {G.}~\bibnamefont
  {Stupakov}},\ }\href@noop {} {\bibfield  {journal} {\bibinfo  {journal}
  {Nuclear Instruments and Methods in Physics Research Section A: Accelerators,
  Spectrometers, Detectors and Associated Equipment}\ }\textbf {\bibinfo
  {volume} {690}},\ \bibinfo {pages} {106} (\bibinfo {year}
  {2012})}\BibitemShut {NoStop}%
\bibitem [{\citenamefont {Antipov}\ \emph {et~al.}(2012)\citenamefont
  {Antipov}, \citenamefont {Jing}, \citenamefont {Fedurin}, \citenamefont
  {Gai}, \citenamefont {Kanareykin}, \citenamefont {Kusche}, \citenamefont
  {Schoessow}, \citenamefont {Yakimenko},\ and\ \citenamefont
  {Zholents}}]{antipov2012experimental}%
  \BibitemOpen
  \bibfield  {author} {\bibinfo {author} {\bibfnamefont {S.}~\bibnamefont
  {Antipov}}, \bibinfo {author} {\bibfnamefont {C.}~\bibnamefont {Jing}},
  \bibinfo {author} {\bibfnamefont {M.}~\bibnamefont {Fedurin}}, \bibinfo
  {author} {\bibfnamefont {W.}~\bibnamefont {Gai}}, \bibinfo {author}
  {\bibfnamefont {A.}~\bibnamefont {Kanareykin}}, \bibinfo {author}
  {\bibfnamefont {K.}~\bibnamefont {Kusche}}, \bibinfo {author} {\bibfnamefont
  {P.}~\bibnamefont {Schoessow}}, \bibinfo {author} {\bibfnamefont
  {V.}~\bibnamefont {Yakimenko}}, \ and\ \bibinfo {author} {\bibfnamefont
  {A.}~\bibnamefont {Zholents}},\ }\href@noop {} {\bibfield  {journal}
  {\bibinfo  {journal} {Physical Review Letters}\ }\textbf {\bibinfo {volume}
  {108}},\ \bibinfo {pages} {144801} (\bibinfo {year} {2012})}\BibitemShut
  {NoStop}%
\bibitem [{\citenamefont {Deng}\ \emph {et~al.}(2014)\citenamefont {Deng},
  \citenamefont {Zhang}, \citenamefont {Feng}, \citenamefont {Zhang},
  \citenamefont {Wang}, \citenamefont {Lan}, \citenamefont {Feng},
  \citenamefont {Zhang}, \citenamefont {Liu}, \citenamefont {Yao} \emph
  {et~al.}}]{deng2014experimental}%
  \BibitemOpen
  \bibfield  {author} {\bibinfo {author} {\bibfnamefont {H.}~\bibnamefont
  {Deng}}, \bibinfo {author} {\bibfnamefont {M.}~\bibnamefont {Zhang}},
  \bibinfo {author} {\bibfnamefont {C.}~\bibnamefont {Feng}}, \bibinfo {author}
  {\bibfnamefont {T.}~\bibnamefont {Zhang}}, \bibinfo {author} {\bibfnamefont
  {X.}~\bibnamefont {Wang}}, \bibinfo {author} {\bibfnamefont {T.}~\bibnamefont
  {Lan}}, \bibinfo {author} {\bibfnamefont {L.}~\bibnamefont {Feng}}, \bibinfo
  {author} {\bibfnamefont {W.}~\bibnamefont {Zhang}}, \bibinfo {author}
  {\bibfnamefont {X.}~\bibnamefont {Liu}}, \bibinfo {author} {\bibfnamefont
  {H.}~\bibnamefont {Yao}},  \emph {et~al.},\ }\href@noop {} {\bibfield
  {journal} {\bibinfo  {journal} {Physical Review Letters}\ }\textbf {\bibinfo
  {volume} {113}},\ \bibinfo {pages} {254802} (\bibinfo {year}
  {2014})}\BibitemShut {NoStop}%
\end{thebibliography}%

\end{document}